\documentclass[runningheads]{llncs}
\usepackage{times}
\usepackage{epsfig}
\usepackage{graphicx}
\usepackage{comment}
\usepackage{amsmath,amssymb} 
\usepackage{color}
\usepackage{subfigure}

\usepackage[pagebackref=true,breaklinks=true,letterpaper=true,colorlinks,bookmarks=false]{hyperref}

\newcommand{\clP}{\mathcal{P}}
\newcommand{\clX}{\mathcal{X}}
\newcommand{\clY}{\mathcal{Y}}
\newcommand{\clZ}{\mathcal{Z}}
\newcommand{\bbE}{\mathbb{E}}
\newcommand{\bbR}{\mathbb{R}}
\newcommand{\Lrecon}{L_\mathrm{recon}}
\newcommand{\Lguide}{L_\mathrm{guide}}
\newcommand{\Ldistr}{L_\mathrm{distr}}
\newcommand{\Lpdistr}{L'_\mathrm{distr}}
\newcommand{\Lpercp}{L_\mathrm{percp}}
\newcommand{\yguide}{y_\mathrm{guide}}
\newcommand{\KL}{\mathrm{KL}}
\newcommand{\etal}{\textit{et al.}}

\usepackage{ulem}

\begin{document}
\pagestyle{headings}
\mainmatter

\title{Invertible Image Rescaling} 


\titlerunning{Invertible Image Rescaling}
%

\author{Mingqing Xiao\inst{1\footnotemark[1]} \and
Shuxin Zheng\inst{2} \and
Chang Liu\inst{2} \and
Yaolong Wang\inst{3\footnotemark[1]} \and
Di He\inst{1} \and
Guolin Ke\inst{2} \and
Jiang Bian\inst{2} \and
Zhouchen Lin\inst{1} \and
Tie-Yan Liu\inst{2}
}
%
\authorrunning{M. Xiao et al.}
%
\institute{Peking University \and
Microsoft Research Asia \and
Toronto University\\
\email{\{mingqing\_xiao, di\_he, zlin\}@pku.edu.cn, \{shuz, changliu, guoke, jiabia, tyliu\}@microsoft.com, yaolong.wang@mail.utoronto.ca}}
\maketitle

\renewcommand{\thefootnote}{\fnsymbol{footnote}}
\footnotetext[1]{Work done during an internship at Microsoft Research Asia.}

\begin{abstract}

High-resolution digital images are usually downscaled to fit various display screens or save the cost of storage and bandwidth, meanwhile the post-upscaling is adpoted to recover the original resolutions or the details in the zoom-in images. However, typical image downscaling is a non-injective mapping due to the loss of high-frequency information, which leads to the ill-posed problem of the inverse upscaling procedure and poses great challenges for recovering details from the downscaled low-resolution images. Simply upscaling with image super-resolution methods results in unsatisfactory recovering performance. In this work, we propose to solve this problem by modeling the downscaling and upscaling processes from a new perspective, i.e. an invertible bijective transformation, which can largely mitigate the ill-posed nature of image upscaling. We develop an Invertible Rescaling Net (IRN) with deliberately designed framework and objectives to produce visually-pleasing low-resolution images and meanwhile capture the distribution of the lost information using a latent variable following a specified distribution in the downscaling process. In this way, upscaling is made tractable by inversely passing a randomly-drawn latent variable with the low-resolution image through the network. Experimental results demonstrate the significant improvement of our model over existing methods in terms of both quantitative and qualitative evaluations of image upscaling reconstruction from downscaled images.

\end{abstract}

\section{Introduction}
\vspace{-2pt}

With exploding amounts of high-resolution (HR) images/videos on the Internet, image downscaling is quite indispensable for storing, transferring and sharing such large-sized data, as the downscaled counterpart can significantly save the storage, efficiently utilize the bandwidth~\cite{bruckstein2003down,lin2006adaptive,wu2009low,shen2011down,li2018learning} and easily fit for screens with different resolution while maintaining visually valid information~\cite{kim2018task,sun2020learned}.
Meanwhile, many of these downscaling scenarios inevitably raise a great demand for the inverse task, i.e., upscaling the downscaled image to a higher resolution or its original size~\cite{yeo2017will,yeo2018neural,schulter2015fast,giachetti2011real}. However, details are lost and distortions appear when users zoom in or upscale the low-resolution (LR) images. Such an upscaling task is quite challenging since image downscaling is well-known as a non-injective mapping, meaning that there could exist multiple possible HR images resulting in the same downscaled LR image. Hence, this inverse task is usually considered to be ill-posed~\cite{irani2009super,yang2010image,dong2015image}. 

Many efforts have been made to mitigate this ill-posed problem, but the gains fail to meet the expectation. For example, most of previous works choose super-resolution (SR) methods to upscale the downscaled LR images. However, mainstream SR algorithms~\cite{dong2015image,lim2017enhanced,zhang2018residual,zhang2018image,dai2019second,wang2018esrgan} focus only on recovering HR images from LR ones under the guidance of a predefined and non-adjustable downscaling kernel (e.g., Bicubic interpolation), which omits its compatibility to the downscaling operation. Intuitively, as long as the target LR image is pre-downscaled from an HR image, taking the image downscaling method into consideration would be quite invaluable for recovering the high-quality upscaled image.

Instead of simply treating the image downscaling and upscaling as two separate and independent tasks, most recently, there have been efforts~\cite{kim2018task,li2018learning,sun2020learned} attempting to model image downscaling and upscaling as a united task by an encoder-decoder framework. 
Specifically, they proposed to use an upscaling-optimal downscaling method as an encoder which is jointly trained with an upscaling decoder~\cite{kim2018task} or existing SR modules~\cite{li2018learning,sun2020learned}. Although such an integrated training approach can significantly improve the quality of the HR images recovered from the corresponding downscaled LR images, neither can we do a perfect reconstruction. These efforts didn't tackle much on the ill-posedness since they link the two processes only through the training objectives and conduct no attempt to capture any feature of the lost information.

In this paper, with inspiration from the reciprocal nature of this pair of image rescaling tasks, we propose a novel method to largely mitigate this ill-posed problem of the image upscaling. 
According to the Nyquist-Shannon sampling theorem, high-frequency contents are lost during downscaling. Ideally, we hope to keep all lost information to perfectly recover the original HR image, but storing or transferring the high-frequency information is unacceptable.
In order to well address this challenge, we develop a novel invertible model called Invertible Rescaling Net (IRN) which captures some knowledge on the lost information in the form of its distribution and embeds it into model's parameters to mitigate the ill-posedness. Given an HR image $x$, IRN not only downscales it into a visually-pleasing LR image $y$, but also embed the case-specific high-frequency content into an auxiliary case-agnostic latent variable $z$, whose marginal distribution obeys a fixed pre-specified distribution (e.g., isotropic Gaussian). 
Based on this model, we use a randomly drawn sample of $z$ from the pre-specified distribution for the inverse upscaling procedure , which holds the most information that one could have in upscaling.

Yet, there are still several great challenges needed to be addressed during the IRN training process. Specifically, it is essential to ensure the quality of reconstructed HR images, obtain visually pleasing downscaled LR ones, and accomplish the upscaling with a case-agnostic $z$, i.e., $z\sim p(z)$ instead of a case-specific $z \sim p(z|y)$. To this end, we design a novel compact and effective objective function by combining three respective components: an HR reconstruction loss, an LR guidance loss and a distribution matching loss. The last component is for the model to capture the true HR image manifold as well as for enforcing $z$ to be case-agnostic. Neither the conventional adversarial training techniques of generative adversarial nets (GANs)~\cite{goodfellow2014generative} nor the maximum likelihood estimation (MLE) method for existing invertible neural networks~\cite{dinh2015nice,dinh2017density,kingma2018glow,ardizzone2019guided} could achieve our goal, since the model distribution doesn't exist here, meanwhile these methods don't guide the distribution in the latent space.
Instead, we take the pushed-forward empirical distribution of $x$ as the distribution on $y$, which, in independent company with $p(z)$, is the actually used distribution to inversely pass our model to recover the distribution of $x$. We thus match this distribution with the empirical distribution of $x$ (the data distribution). Moreover, due to the invertible nature of our model, we show that once this matching task is accomplished, the matching task in the $(y,z)$ space is also solved, and $z$ is made case-agnostic. We minimize the JS divergence to match the distributions, since the alternative sample-based maximum mean discrepancy (MMD) method~\cite{ardizzone2019analyzing} doesn't generalize well to the high dimension data in our task.

Our contributions are concluded as follows:
\vspace{-2pt}
\begin{itemize}
    \item To our best knowledge, the proposed IRN is the first attempt to model image downscaling and upscaling, a pair of mutually-inverse tasks, using an invertible (i.e., bijective) transformation. Powered by the deliberately designed invertibility, our proposed IRN can largely mitigate the ill-posed nature of image upscaling reconstruction from the downscaled LR image.
    \vspace{-2pt}
    \item We propose a novel model design and efficient training objectives for IRN to enforce the latent variable $z$, with embedded lost high-frequency information in the downscaling direction, to obey a simple case-agnostic distribution. This enables efficient upscaling based on the valuable samples of $z$ drawn from the certain distribution.
    \vspace{-2pt}
    \item The proposed IRN can significantly boost the performance of upscaling reconstruction from downscaled LR images compared with state-of-the-art downscaling-SR and encoder-decoder methods.
    Moreover, the amount of parameters of IRN is significantly reduced, which indicates the light-weight and high-efficiency of the new IRN model.
\end{itemize}

\vspace{-2pt}
\section{Related Work}
\vspace{-4pt}
\subsection{Image Upscaling after Downscaling}
\vspace{-2pt}

Super resolution (SR) is a widely-used image upscaling method and get promising results in low-resolution (LR) image upscaling task. Therefore, SR methods could be used to upscale downscaled images.
Since the SR task is inherently ill-posed, previous SR works mainly focus on learning strong prior information by example-based strategy~\cite{freedman2011image,glasner2009super,schulter2015fast,kim2010single} or deep learning models~\cite{dong2015image,lim2017enhanced,zhang2018residual,zhang2018image,dai2019second,wang2018esrgan}. However, if the targeted LR image is pre-downscaled from the corresponding high-resolution image, taking the image downscaling method into consideration would significantly help the upscaling reconstruction.

Traditional image downscaling approaches employ frequency-based kernels, such as Bilinear, Bicubic, etc.~\cite{mitchell1988reconstruction}, as a low-pass filter to sub-sample the input HR images into target resolution. Normally, these methods suffer from resulting over-smoothed images since the high-frequency details are suppressed. Therefore, several detail-preserving or structurally similar downscaling methods ~\cite{kopf2013content,oeztireli2015perceptually,wang2004image,weber2016rapid,liu2017l_} are proposed recently. Besides those perceptual-oriented downscaling methods, inspired by the potentially mutual reinforcement between downscaling and its inverse task, upscaling, increasing efforts have been focused on the upscaling-optimal downscaling methods, which aim to learn a downscaling model that is optimal to the post-upscaling operation. For instance, Kim \etal~\cite{kim2018task} proposed a task-aware downscaling model based on an auto-encoder framework, in which the encoder and decoder act as the downscaling and upscaling model, respectively, such that the downscaling and upscaling processes are trained jointly as a united task. Similarly, Li \etal~\cite{li2018learning} proposed to use a CNN to estimate downscaled compact-resolution images and leverage a learned or specified SR model for HR image reconstruction. More recently, Sun \etal~\cite{sun2020learned} proposed a new content-adaptive-resampler based image downscaling method, which can be jointly trained with any existing differentiable upscaling (SR) models. 
Although these attempts have an effect of pushing one of downscaling and upscaling to resemble the inverse process of the other, they still suffer from the ill-posed nature of image upscaling problem. 
In this paper, we propose to model the downscaling and upscaling processes by leveraging the invertible neural networks.

\textbf{Difference from SR.} Note that image upscaling is a different task from super-resolution. In our scenario, the ground-truth HR image is available at the beginning but somehow we have to discard it and store/transmit the LR version instead. We hope that we can recover the HR image afterwards using the LR image. While for SR, the real HR is unavailable in applications and the task is to generate new HR images for LR ones. 

\vspace{-10pt}
\subsection{Invertible Neural Network}
\vspace{-2pt}

The invertible neural network (INN)~\cite{dinh2015nice,dinh2017density,kingma2018glow,kumar2019videoflow,grathwohl2019ffjord,behrmann2019invertible,chen2019residual} is a popular choice for generative models, in which the generative process $x = f_\theta (z)$ given a latent variable $z$ can be specified by an INN architecture $f_\theta$. The direct access to the inverse mapping $z = f_\theta^{-1} (x)$ makes inference much cheaper. As it is possible to compute the density of the model distribution in INN explicitly, one can use the maximum likelihood method for training. Due to such flexibility, INN architectures are also used for many variational inference tasks~\cite{rezende2015variational,kingma2016improved,berg2018sylvester}.

INN is composed of invertible blocks. In this study, we employ the invertible architecture in~\cite{dinh2017density}. For the $l$-th block, input $h^l$ is split into $h_1^l$ and $h_2^l$ along the channel axis, and they undergo the additive affine transformations~\cite{dinh2015nice}:
\begin{eqnarray}
    \begin{aligned}
        &h_1^{l+1} = h_1^l + \phi(h_2^l),\\
        &h_2^{l+1} = h_2^l + \eta(h_1^{l+1}),
    \end{aligned}
    \label{eq:invblock}
\end{eqnarray}
where $\phi, \eta$ are arbitrary functions. The corresponding output is [$h_1^{l+1}, h_2^{l+1}$]. Given the output, its inverse transformation is easily computed:
\begin{eqnarray}
    \begin{aligned}
        &h_2^l = h_2^{l+1} - \eta(h_1^{l+1}),\\
        &h_1^l = h_1^{l+1} - \phi(h_2^l),
    \end{aligned}
\end{eqnarray}
To enhance the transformation ability, the identity branch is often augmented~\cite{dinh2017density}:
\begin{eqnarray}
    \begin{aligned}
        &h_1^{l+1} = h_1^l \odot \exp(\psi(h_2^l)) + \phi(h_2^l),\\
        &h_2^{l+1} = h_2^l \odot \exp(\rho(h_1^{l+1})) + \eta(h_1^{l+1}),\\
        &h_2^l = (h_2^{l+1} - \eta(h_1^{l+1})) \odot \exp(-\rho(h_1^{l+1})),\\
        &h_1^l = (h_1^{l+1} - \phi(h_2^l)) \odot \exp(-\psi(h_2^l)).
    \end{aligned}
    \label{eq:invblockexp}
\end{eqnarray}

Some prior works studied using INN for paired data $(x,y)$. Ardizzone \etal~\cite{ardizzone2019analyzing} analyzed real-world problems from medicine and astrophysics. Compared to their tasks, image downscaling and upscaling bring more difficulties because of notably larger dimensionality, so that their losses do not work for our task. 
In addition, the ground-truth LR image $y$ does not exist in our task.
Guided image generation and colorization using INN is proposed in~\cite{ardizzone2019guided} where the invertible modeling between $x$ and $z$ is conditioned on a guidance $y$. The model cannot generate $y$ given $x$ thus is unsuitable for the image upscaling task. INN is also applied to the image-to-image translation task~\cite{van2019reversible} where the paired domain $(X,Y)$ instead of paired data is considered, thus is again not the case of image upscaling.

\subsection{Image Compression}
Image compression is a type of data compression applied to digital images, to reduce their cost for storage or transmission. Image compression may be lossy (e.g., JPEG, BPG) or lossless (e.g., PNG, BMP). Recently, deep learning based image compression methods~\cite{balle2016end,rippel2017real,balle2018variational,agustsson2019generative,minnen2018joint} show promising results on both visual effect and compression ratio. However, the resolution of image won't be changed by compression, which means there is no visually meaningful low-resolution image but only bit-stream after compressing. Thus our task can't be served by image compression methods.

\section{Methods}

\subsection{Model Specification}\label{sec:model-spec}


\begin{figure} [ht]
    \centering
    \includegraphics[scale=0.45]{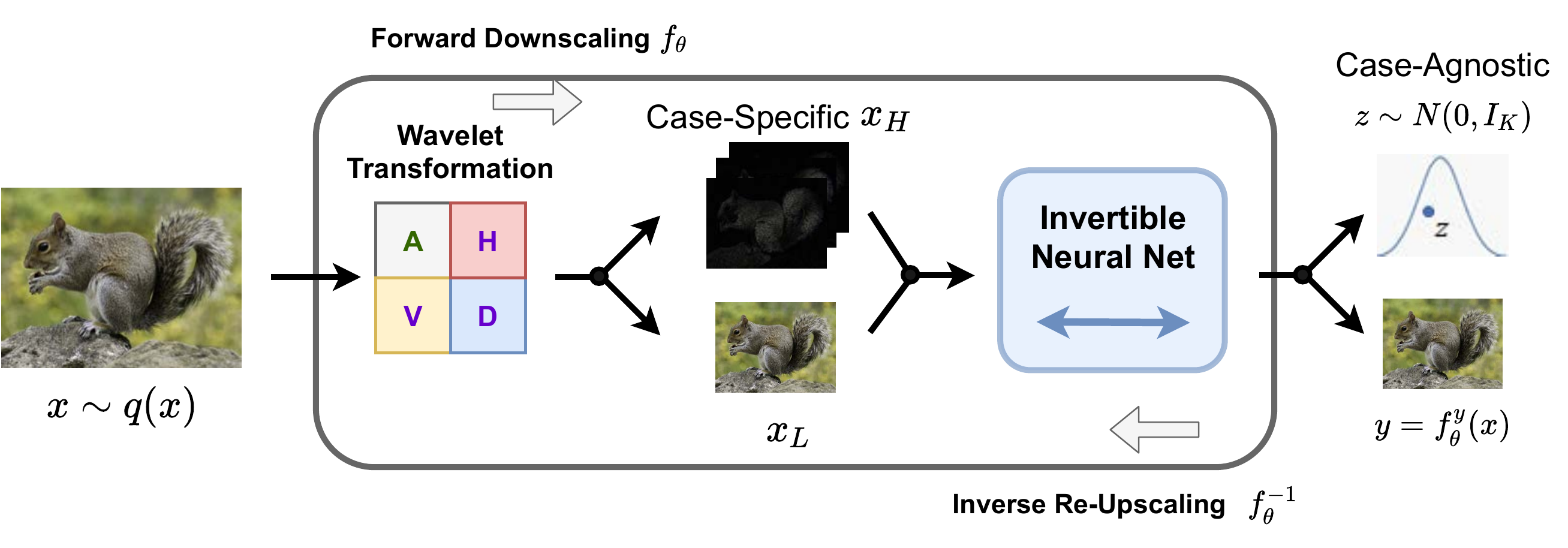}
    \caption{Illustration of the problem formulation. In the forward downscaling procedure, HR image $x$ is transformed to visually pleasing LR image $y$ and case-agnostic latent variable $z$ through a parameterized invertible function $f_{\theta}(\cdot)$; in the inverse upscaling procedure, a randomly drawn $z$ combined with LR image $y$ are transformed to HR image through the inverse function $f_{\theta}^{-1}(\cdot)$.}
    \label{fig:problem formulation}
    \vspace{-10pt}
\end{figure}

The sketch of our modeling framework is presented in Fig.~\ref{fig:problem formulation}. As explained in Introduction, we mitigate the ill-posed problem of the upscaling task by modeling the distribution of lost information during downscaling. We note that according to the Nyquist-Shannon sampling theorem~\cite{shannon1949communication}, the lost information during downscaling an HR image amounts to high-frequency contents.
Thus we firstly employ a wavelet transformation to decompose the HR image $x$ into low and high-frequency component, denote as $x_L$ and $x_H$ respectively. Since the case-specific high-frequency information will be lost after downscaling, in order to best recover the original $x$ as possible in the upscaling procedure, we use an invertible neural network to produce the visually-pleasing LR image $y$ meanwhile model the distribution of the lost information by introducing an auxiliary latent variable $z$. In contrast to the case-specific $x_H$ (i.e., $x_H \sim p(x_H | x_L)$), we force $z$ to be case-agnostic (i.e., $z \sim p(z)$) and obey a simple specified distribution, e.g., an isotropic Gaussian distribution. In this way, there is no further need to preserve either $x_H$ or $z$ after downscaling, and $z$ can be randomly sampled in the upscaling procedure, which is used to reconstruct $x$ combined with LR image $y$ by inversely passing the model.

\subsection{Invertible Architecture}
\vspace{-10pt}
\begin{figure*}[ht]
	\centering
	\includegraphics[width=\textwidth]{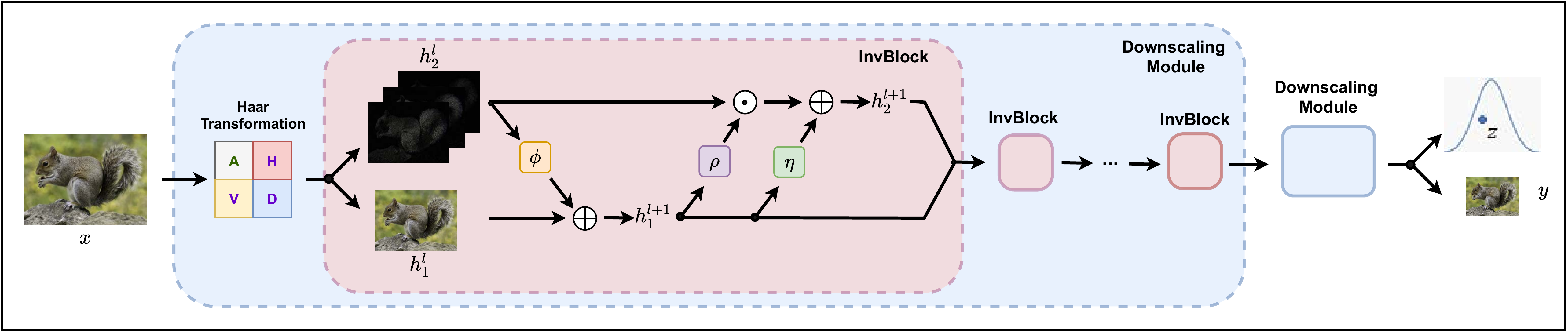}
	\caption{Illustration of our framework. The invertible architecture is composed of Downscaling Modules, in which InvBlocks are stacked after a Haar Transformation. Each Downscaling Module reduces the spatial resolution by 2$\times$. The $\exp(\cdot)$ of $\rho$ is omit.}
	\label{fig:framework}
	\vspace{-10pt}
\end{figure*}

The general architecture of our proposed IRN is composed of stacked \textit{Downscaling Modules}, each of which contains one \textit{Haar Transformation} block and several invertible neural network blocks (\textit{InvBlocks}), as illustrated in Fig.~\ref{fig:framework}. We will show later that both of them are invertible, and thus the entire IRN model is invertible accordingly. 

\textbf{The Haar Transformation}\quad We design the model to contain certain inductive bias, which can efficiently learn to decompose $x$ into the downscaled image $y$ and case-agnostic high-frequency information embedded in $z$. To achieve this, we apply the Haar Transformation as the first layer in each downscaling module, which can explicitly decompose the input images into an approximate low-pass representation, and three directions of high-frequency coefficients \cite{wilson2006facial}\cite{lienhart2002extended}\cite{ardizzone2019guided}. More concretely, the Haar Transformation transforms the input raw images or a group of feature maps with height $H$, width $W$ and channel $C$ into a tensor of shape $(\frac{1}{2}H, \frac{1}{2}W, 4C)$. The first $C$ slices of the output tensor are effectively produced by an average pooling, which is approximately a low-pass representation equivalent to the Bilinear interpolation downsampling. The rest three groups of $C$ slices contain residual components in the vertical, horizontal and diagonal directions respectively, which are the high-frequency information in the original HR image. By such a transformation, the low and high-frequency information are effectively separated and will be fed into the following InvBlocks.

\textbf{InvBlock}\quad Taking the feature maps after the Haar Transformation as input, a stack of InvBlocks is used to further abstract the LR and latent representations. We leverage the general coupling layer architecture proposed in~\cite{dinh2015nice,dinh2017density}, i.e. Eqs.~(\ref{eq:invblock},\ref{eq:invblockexp}). 

Utilizing the coupling layer is based on our considerations that (1) the input has already been split into low and high-frequency components by the Haar transformation; (2) we want the two branches of the output of a coupling layer to further polish the low and high-frequency inputs for a suitable LR image appearance and an independent and properly distributed latent representation of the high-frequency contents.
So we match the low and high-frequency components respectively to the split of $h_1^l$, $h_2^l$ in Eq.~\eqref{eq:invblock}. 
Furthermore, as the shortcut connection is proved to be important in the image scaling tasks~\cite{lim2017enhanced,wang2018esrgan}, we employ the additive transformation (Eq.~\ref{eq:invblock}) for the low-frequency part $h_1^l$, and the enhanced affine transformation (Eq.~\ref{eq:invblockexp}) for the high-frequency part $h_2^l$ to increase the model capacity, as shown in Fig.~\ref{fig:framework}.

Note that the transformation functions $\phi(\cdot), \eta(\cdot), \rho(\cdot)$ in Fig.~\ref{fig:framework} can be arbitrary. Here we employ a densely connected convolutional block, which is referred as Dense Block in~\cite{wang2018esrgan} and demonstrated for its effectiveness of image upscaling task. Function $\rho(\cdot)$ is further followed by a centered sigmoid function and a scale term to prevent numerical explosion due to the $\exp(\cdot)$ function. Note that Figure~\ref{fig:framework} omits the $\exp(\cdot)$ in function $\rho$.

\textbf{Quantization}\quad To save the output images of IRN as common image storage format such as RGB (8 bits for each R, G and B color channels), a quantization module is adopted which converts floating-point values of produced LR images to 8-bit unsigned int. We simply use rounding operation as the quantization module, store our output LR images by PNG format and use it in the upscaling procedure. There is one obstacle should be noted that the quantization module is nondifferentiable. To ensure that IRN can be optimized during training, we use Straight-Through Estimator \cite{bengio2013estimating} on the quantization module when calculating the gradients.

\subsection{Training Objectives}\label{sec:train-method}

Based on Section~\ref{sec:model-spec}, our approach for invertible downscaling constructs a model that specifies a correspondence between HR image $x$ and LR image $y$, as well as a case-agnostic distribution $p(z)$ of $z$.
The goal of training is to drive these modeled relations and quantities to match our desiderata and HR image data $\{x^{(n)}\}_{n=1}^N$.
This includes three specific goals, as detailed below.

{\textbf{LR Guidance}}\quad
Although the invertible downscaling task does not pose direct requirements on the produced LR images, we do hope that they are valid visually pleasing LR images.
To achieve this, we utilize the widely acknowledged Bicubic method~\cite{mitchell1988reconstruction} to guide the downscaling process of our model.
Let $\yguide^{(n)}$ be the LR image corresponding to $x^{(n)}$ that is produced by the Bicubic method.
To make our model follow the guidance, we drive the model-produced LR image $f_\theta^y (x^{(n)})$ to resemble $\yguide^{(n)}$:
\begin{align}
    \Lguide(\theta) := \sum_{n=1}^N \ell_\clY (\yguide^{(n)}, f_\theta^y (x^{(n)})),
    \label{eq.guide loss}
\end{align}
where $\ell_\clY$ is a difference metric on $\clY$, e.g., the $L_1$ or $L_2$ loss.
We call it the LR guidance loss.
This practice has also been adopted in the literature~\cite{kim2018task,sun2020learned}.

{\textbf{HR Reconstruction}}\quad
Although $f_\theta$ is invertible, it is not for the correspondence between $x$ and $y$ when $z$ is not transmitted.
We hope that for a specific downscaled LR image $y$, the original HR image can be restored by the model using any sample of $z$ from the case-agnostic $p(z)$.
Inversely, this also encourages the forward process to produce a disentangled representation of $z$ from $y$.
As described in Section~\ref{sec:model-spec}, given a HR image $x^{(n)}$, the model-downscaled LR image $f_\theta^y (x^{(n)})$ is to be upscaled by the model as $f_\theta^{-1} (f_\theta^y (x^{(n)}), z)$ with a randomly drawn $z \sim p(z)$.
The reconstructed HR image should match the original one $x^{(n)}$, so we minimize the expected difference and traverse over all the HR images:
\begin{align}
    \Lrecon(\theta) := \sum_{n=1}^N \bbE_{p(z)} [ \ell_\clX ( x^{(n)}, f_\theta^{-1} (f_\theta^y (x^{(n)}), z) ) ],
    \label{eq.recon loss}
\end{align}
where $\ell_\clX$ measures the difference between the original image and the reconstructed one.
We call $\Lrecon(\theta)$ the HR reconstruction loss.
For practical minimization, we estimate the expectation w.r.t. $z$ by one random draw from $p(z)$ for each evaluation.

{\textbf{Distribution Matching}}\quad
The third part of the training goal is to encourage the model to catch the data distribution $q(x)$ of HR images, demonstrated by its sample cloud $\{x^{(n)}\}_{n=1}^N$.
Recall that the model reconstructs a HR image $x^{(n)}$ by $f_\theta^{-1} (y^{(n)}, z^{(n)})$, where $y^{(n)} := f_\theta^y (x^{(n)})$ is the model-downscaled LR image, and $z^{(n)} \sim p(z)$ is the randomly drawn latent variable.
When traversing over the sample cloud of true HR images $\{x^{(n)}\}_{n=1}^N$, $\{y^{(n)}\}_{n=1}^N$ also form a sample cloud of a distribution.
We denote this distribution with the push-forward notation as ${f_\theta^y}_\# [q(x)]$, which represents the distribution of the transformed random variable $f_\theta^y (x)$ where the original random variable $x$ obeys distribution $q(x)$, $x \sim q(x)$.
Similarly, the sample cloud $\{f_\theta^{-1} (y^{(n)}, z^{(n)})\}_{n=1}^N$ represents the distribution of model-reconstructed HR images, and we denote it as ${f_\theta^{-1}}_\# \big[ {f_\theta^y}_\# [q(x)] \; p(z) \big]$ since $(y^{(n)}, z^{(n)}) \sim {f_\theta^y}_\# [q(x)] \; p(z)$ (note that $y^{(n)}$ and $z^{(n)}$ are independent due to the generation process).
The desideratum of distribution matching is to drive the model-reconstructed distribution towards data distribution, which can be achieved by minimizing their difference measured by some metric of distributions:
\begin{align}
    \Ldistr(\theta) := L_\clP \big( {f_\theta^{-1}}_\# \big[ {f_\theta^y}_\# [q(x)] \; p(z) \big], q(x) \big).
    \label{eq.distr loss}
\end{align}

The distribution matching loss pushes the model-reconstructed HR images to lie on the manifold of true HR images so as to make the recovered images appear more realistic.
It also drives the case-independence of $z$ from $y$ in the forward process.
To see this, we note that if $f_\theta$ is invertible, then in the asymptotic case, the two distributions match on $\clX$, i.e., ${f_\theta^{-1}}_\# \big[ {f_\theta^y}_\# [q(x)] \; p(z) \big] = q(x)$, if and only if they match on $\clY \times \clZ$, i.e., ${f_\theta^y}_\# [q(x)] \; p(z) = {f_\theta}_\# [q(x)]$.
The loss thus drives the coupled distribution ${f_\theta}_\# [q(x)] = (f_\theta^y, f_\theta^z)_\# [q(x)]$ of $(y, z)$ from the forward process towards the decoupled distribution ${f_\theta^y}_\# [q(x)] \; p(z)$.
Neither effect can be fully guaranteed by the reconstruction and guidance losses.

As mentioned in Introduction, the minimization is generally hard since both distributions are high-dimensional and have unknown density function.
We employ the JS divergence as the probability metric $L_\clP$, and our distribution matching loss can be estimated in the following way:
\begin{align}
	& \Ldistr(\theta) = \mathrm{JS}( {f_\theta^{-1}}_\# \big[ {f_\theta^y}_\# [q(x)] \; p(z) \big], q(x) ) \notag \\
	\approx{} & \frac{1}{2 N} \max_T \sum_n \Big\{ \log \sigma(T (x^{(n)})) \notag \\
	    & {} + \log \Big( \! 1 - \sigma \big[ T \big( f_\theta^{-1} ( f_\theta^y (x^{(n)}), z^{(n)} ) \big) \big] \! \Big) \! \Big\} + \log 2,
	\label{eqn:bwddismat}
\end{align}
where $\{z^{(n)}\}_{n=1}^N$ are i.i.d. samples from $p(z)$, $\sigma$ is the sigmoid function, $T: \clX \to \bbR$ is a function on $\clX$ ($\sigma(T(\cdot))$ is regarded as a discriminator in GAN literatures), and ``$\approx$'' is due to Monte Carlo estimation.
The appendix provides the details.
For practical computation, the function $T$ is parameterized as a neural network $T_\phi$ and $\max_T$ amounts to $\max_\phi$.
The expression~\eqref{eqn:bwddismat} is also suitable for estimating its gradient w.r.t. $\theta$ and $\phi$, thus optimization is made practical.

{\textbf{Total Loss}}\quad
We optimize our IRN model by minimizing the compact loss $L_\mathrm{total}(\theta)$ with the combination of HR reconstruction loss $\Lrecon(\theta)$, LR guidance loss $\Lguide(\theta)$ and distribution matching loss $\Ldistr(\theta)$:
\begin{align}
    L_\mathrm{total} := \lambda_1 \Lrecon + \lambda_2 \Lguide + \lambda_3 \Ldistr,
\end{align}
where $\lambda_1, \lambda_2, \lambda_3$ are coefficients for balancing different loss terms.

{\textbf{Loss Minimization in Practice}}\quad
As an issue in practice, we find that directly minimizing the total loss $L_\mathrm{total}(\theta)$ is difficult to train, due to the unstable training process of GANs~\cite{arjovsky2017towards}.
We propose a pre-training stage that adopts a weakened but more stable surrogate of the distribution matching loss.
Recall that the distribution matching loss $L_\clP \big( {f_\theta^{-1}}_\# \big[ {f_\theta^y}_\# [q(x)] \; p(z) \big], q(x) \big)$ on $\clX$ has the same asymptotic effect as the loss $L_\clP( {f_\theta^y}_\# [q(x)] \; p(z), (f_\theta^y, f_\theta^z)_\# [q(x)] )$ on $\clY \times \clZ$.
The surrogate considers partial distribution matching on $\clZ$, i.e., $L_\clP( p(z), {f_\theta^z}_\# [q(x)] )$.
Since the density function of one of the distributions, $p(z)$, is now made available, we can choose more stable distribution metrics for minimization, such as the cross entropy (CE):
\begin{align}
    & \Lpdistr(\theta) := \mathrm{CE}( {f_\theta^z}_\# [q(x)], p(z) ) \notag \\
    =& - \! \bbE_{{f_\theta^z}_\# [q(x)]} [\log p(z)] = - \bbE_{q(x)} [\log p( z \!=\! f_\theta^z (x) )].
    \label{eq.distr loss'}
\end{align}
A related training method is the maximum likelihood estimation (MLE), i.e., \newline
$\max_\theta \bbE_{q(x)} [ \log {f_\theta^{-1}}_\# [p(y, z)] ]$,
which is widely adopted by prevalent flow-based generative models~\cite{dinh2015nice,dinh2017density,kingma2018glow,ardizzone2019guided}.
It is equivalent to minimizing the Kullback-Leibler (KL) divergence $\KL( q(x), {f_\theta^{-1}}_\# [p(y, z)] )$.
The mentioned models explicitly specify the density function of $p(y, z)$, thus the density function of ${f_\theta^{-1}}_\# [p(y, z)]$ is made available together with the tractable Jacobian determinant computation of $f_\theta$.
However, the same objective cannot be leveraged for our model since we do not have the density function for ${f_\theta^y}_\# [q(x)] \; p(z)$; only that of $p(z)$ is known\footnote{
    MLEs corresponding to minimizing $\KL( q(x|y), {f_\theta^{-1}(y, \cdot)}_\# [p(z)] )$ or $\KL\Big( q(x), \Big( \bbE_{{f_\theta^y}_\# [q(x)]} [f_\theta^{-1}(y, \cdot)] \Big)_\# [p(z)] \Big)$ are also impossible, since the pushed-forward distributions have a.e. zero density in $\clX$ so the KL is a.e. infinite.
}.
The invertible neural network (INN)~\cite{ardizzone2019analyzing} meets the same problem and cannot use MLE either.

We call IRN as our model trained by minimizing the following total objective:
\begin{align}
    L_\mathrm{IRN} := \lambda_1 \Lrecon + \lambda_2 \Lguide + \lambda_3 \Lpdistr.
    \label{eq.IRN objective function}
\end{align}

After the pre-training stage, we restore the full distribution matching loss $\Ldistr$ in the objective in place of $\Lpdistr$.
Additionally, we also employ a perceptual loss~\cite{johnson2016perceptual} $\Lpercp$ on $\clX$, which measures the difference of two images via their semantic features extracted by benchmarking models.
It enhances the perceptual similarity between generated and true images thus helps to produce more realistic images.
The perceptual loss has several slightly modified variants which mainly differ in the position of the objective features~\cite{ledig2017photo}\cite{wang2018esrgan}.
We adopt the variant proposed in~\cite{wang2018esrgan}.
We call IRN+ as our model trained by minimizing the following total objective:
\begin{align*}
    L_\mathrm{IRN+} := \lambda_1 \Lrecon + \lambda_2 \Lguide + \lambda_3 \Ldistr + \lambda_4 \Lpercp.
\end{align*}

{\textbf{Difference with GAN}}\quad On one hand, although the JS divergence is adopted to instead of MLE as distribution matching loss for optimizing IRN, there is one thing should be noted that our model is totally different from typical GAN models: besides the latent variable $z$ which has a prior, there exists $y$ in IRN model which is subject to some distributional constraints, and our model does not have a standalone distribution on $x$. Therefore, the conventional way to use adversarial loss simply cannot be applied, and we match towards the data distribution with an essentially different distribution from the GAN model distribution. On the other hand, except for JS divergence, a CE loss for $L'_{\mathrm{distr}}$ is also adopted as distribution matching loss of IRN. In general, the distribution matching loss reflects the essential idea of IRN, which is totally different from GAN.

\vspace{-2pt}
\section{Experiments}\label{sec:exp}

\vspace{-4pt}
\subsection{Dataset and Settings}
\vspace{-2pt}

We employ the widely used DIV2K~\cite{agustsson2017ntire} image restoration dataset to train our model, which contains 800 high-quality 2K resolution images in the training set, and 100 in the validation set. Besides, we evaluate our model on 4 additional standard datasets, i.e. the Set5~\cite{bevilacqua2012low}, Set14~\cite{zeyde2010single}, BSD100~\cite{martin2001database}, and Urban100~\cite{huang2015single}. Following the setting in~\cite{lim2017enhanced}, we quantitatively evaluate the peak noise-signal ratio (PSNR) and SSIM~\cite{wang2004image} on the Y channel of images represented in the YCbCr (Y, Cb, Cr) color space. Due to space constraint, we leave training strategy details in the appendix.

\subsection{Evaluation on Reconstructed HR Images}
\vspace{-2pt}

This section reports the quantitative and qualitative performance of HR image reconstruction with different downscaling and upscaling methods. We consider two kinds of reconstruction methods as our baselines: (1) downscaling with Bicubic interpolation and upscaling with state-of-the-art SR models~\cite{dong2015image,lim2017enhanced,zhang2018residual,zhang2018image,wang2018esrgan,dai2019second}; (2) downscaling with upscaling-optimal models~\cite{kim2018task,li2018learning,sun2020learned} and upscaling with SR models. For the method of~\cite{wang2018esrgan}, we denote ESRGAN as their pre-trained model, and ESRGAN+ as their GAN-based model. We further investigate the influence of different $z$ samples on the reconstructed image $x$. Finally, we empirically study the effectiveness of the different types of loss in the pre-training stage.

\begin{table*} [ht]
\vspace{-10pt}
	\centering
	\scriptsize
	\caption{Quantitative evaluation results (PSNR / SSIM) of different downscaling and upscaling methods for image reconstruction on benchmark datasets: Set5, Set14, BSD100, Urban100, and DIV2K validation set. For our method, differences on average PSNR / SSIM from different $z$ samples are less than 0.02. We report the mean result over 5 draws.}
	\begin{tabular}{c|c|c|c|c|c|c|c}
		\hline
		Downscaling \& Upscaling  &  Scale & Param  &  Set5  &  Set14  &  BSD100  &  Urban100  &  DIV2K \\
		
		\hline  
		\hline
		Bicubic \& Bicubic & 2$\times$ & / & 33.66 / 0.9299 & 30.24 / 0.8688 & 29.56 / 0.8431 & 26.88 / 0.8403 & 31.01 / 0.9393 \\
		
		\hline
		Bicubic \& SRCNN \cite{dong2015image} & 2$\times$ & 57.3K &  36.66 / 0.9542  &  32.45 / 0.9067  &  31.36 / 0.8879  &  29.50 / 0.8946 &   --  \\
		
		
		\hline
		Bicubic \& EDSR \cite{lim2017enhanced} & 2$\times$ &  40.7M & 38.20 / 0.9606  &  34.02 / 0.9204  &  32.37 / 0.9018  &  33.10 / 0.9363 &  35.12 / 0.9699 \\
		
		\hline
		Bicubic \& RDN \cite{zhang2018residual} & 2$\times$ & 22.1M &  38.24 / 0.9614  &  34.01 / 0.9212  &  32.34 / 0.9017  &  32.89 / 0.9353 &   --  \\
		
		\hline
		Bicubic \& RCAN \cite{zhang2018image} & 2$\times$ & 15.4M &  38.27 / 0.9614  &  34.12 / 0.9216  &  32.41 / 0.9027  &  33.34 / 0.9384  &  --  \\
		
		\hline
		Bicubic \& SAN \cite{dai2019second} & 2$\times$ & 15.7M &  38.31 / 0.9620  &  34.07 / 0.9213  &  32.42 / 0.9028  &  33.10 / 0.9370  &  --  \\
		
		
		\hline
		TAD \& TAU \cite{kim2018task} & 2$\times$ & -- & 38.46 /  --   & 35.52 /  --  & 36.68 /  --   & 35.03 /  --  & 39.01 /  --  \\
		
		
		\hline
		CNN-CR \& CNN-SR \cite{li2018learning} & 2$\times$ & -- & 38.88 / -- & 35.40 / -- & 33.92 / -- & 33.68 / -- & --\\
		
		\hline
		CAR \& EDSR \cite{sun2020learned} & 2$\times$ & 51.1M & 38.94 / 0.9658 & 35.61 / 0.9404 & 33.83 / 0.9262 & 35.24 / 0.9572 & 38.26 / 0.9599 \\
		
		\hline
		IRN (ours) & 2$\times$ & 1.66M & \textcolor{red}{43.99} / \textcolor{red}{0.9871} & \textcolor{red}{40.79} / \textcolor{red}{0.9778} & \textcolor{red}{41.32} / \textcolor{red}{0.9876} & \textcolor{red}{39.92} / \textcolor{red}{0.9865} & \textcolor{red}{44.32} / \textcolor{red}{0.9908} \\
		
		\hline  
		\hline
		Bicubic \& Bicubic & 4$\times$ & / & 28.42 / 0.8104 & 26.00 / 0.7027 & 25.96 / 0.6675 & 23.14 / 0.6577 & 26.66 / 0.8521 \\
		
		\hline
		Bicubic \& SRCNN \cite{dong2015image} & 4$\times$ & 57.3K &  30.48 / 0.8628  &  27.50 / 0.7513  &  26.90 / 0.7101  &  24.52 / 0.7221 &   --  \\
		
		
		\hline
		Bicubic \& EDSR \cite{lim2017enhanced} & 4$\times$ & 43.1M & 32.62 / 0.8984 & 28.94 / 0.7901 & 27.79 / 0.7437 & 26.86 / 0.8080 & 29.38 / 0.9032 \\

		\hline
		Bicubic \& RDN \cite{zhang2018residual} & 4$\times$ & 22.3M &  32.47 / 0.8990  &  28.81 / 0.7871  &  27.72 / 0.7419  &  26.61 / 0.8028 &   --  \\
		
		\hline
		Bicubic \& RCAN \cite{zhang2018image} & 4$\times$ & 15.6M & 32.63 / 0.9002 & 28.87 / 0.7889 & 27.77 / 0.7436 & 26.82 / 0.8087 & 30.77 / 0.8460 \\
		
		\hline
		Bicubic \& ESRGAN \cite{wang2018esrgan} & 4$\times$ & 16.3M & 32.74 / 0.9012 & 29.00 / 0.7915 & 27.84 / 0.7455 & 27.03 / 0.8152 & 30.92 / 0.8486 \\
		
		\hline
		Bicubic \& SAN \cite{dai2019second} & 4$\times$ & 15.7M &  32.64 / 0.9003  &  28.92 / 0.7888  &  27.78 / 0.7436  &  26.79 / 0.8068  &  --  \\
		
		\hline
		TAD \& TAU \cite{kim2018task} & 4$\times$ & -- & 31.81 /  --  & 28.63 /  --   & 28.51 /  --   & 26.63 /  --   & 31.16 /  --   \\
		
		
		\hline
		CAR \& EDSR \cite{sun2020learned} & 4$\times$ & 52.8M & 33.88 / 0.9174 & 30.31 / 0.8382 & 29.15 / 0.8001 & 29.28 / 0.8711 & 32.82 / 0.8837 \\
		
		\hline
		IRN (ours) & 4$\times$ & 4.35M & \textcolor{red}{36.19} / \textcolor{red}{0.9451} & \textcolor{red}{32.67} / \textcolor{red}{0.9015} & \textcolor{red}{31.64} / \textcolor{red}{0.8826} & \textcolor{red}{31.41} / \textcolor{red}{0.9157} & \textcolor{red}{35.07} / \textcolor{red}{0.9318} \\

        \hline
	    
	\end{tabular}
	
    \label{quantitative results}
    \vspace{-10pt}
\end{table*}

\noindent\textbf{Quantitative Results}\quad
Table~\ref{quantitative results} summarizes the quantitative comparison results of different reconstruction methods where IRN significantly outperforms previous state-of-the-art methods regarding PSNR and SSIM in all datasets. We leave the results of IRN+ in the appendix because it is a visual-perception-oriented model. As shown in Table~\ref{quantitative results}, upscaling-optimal downscaling models largely enhance the reconstruction of HR images by state-of-the-art SR models compared with downscaling with Bicubic interpolation. However, they still hardly achieve satisfying results due to the ill-posed nature of upscaling. In contract, with the invertibility, IRN significantly boosts the PSNR metric about 4-5 dB and 2-3 dB on each benchmark dataset in $2\times$ and $4\times$ scale downsampling and reconstruction, and the improvement goes as large as 5.94 dB compared with the state-of-the-art downscaling and upscaling model. These results indicate an exponential improvement of IRN in the reduction of information loss, which also accords with the significant improvement in SSIM. 

Moreover, the number of parameters of IRN is relatively small. When Bicubic downscaling and super-resolution methods require large model size (\textgreater15M) for better results, our IRN only has 1.66M and 4.35M parameters in scale 2$\times$ and 4$\times$ respectively. It indicates that our model is light-weight and efficient.

\begin{figure*} [ht]
 \vspace{-10pt}
    \centering
    \includegraphics[scale=0.0875]{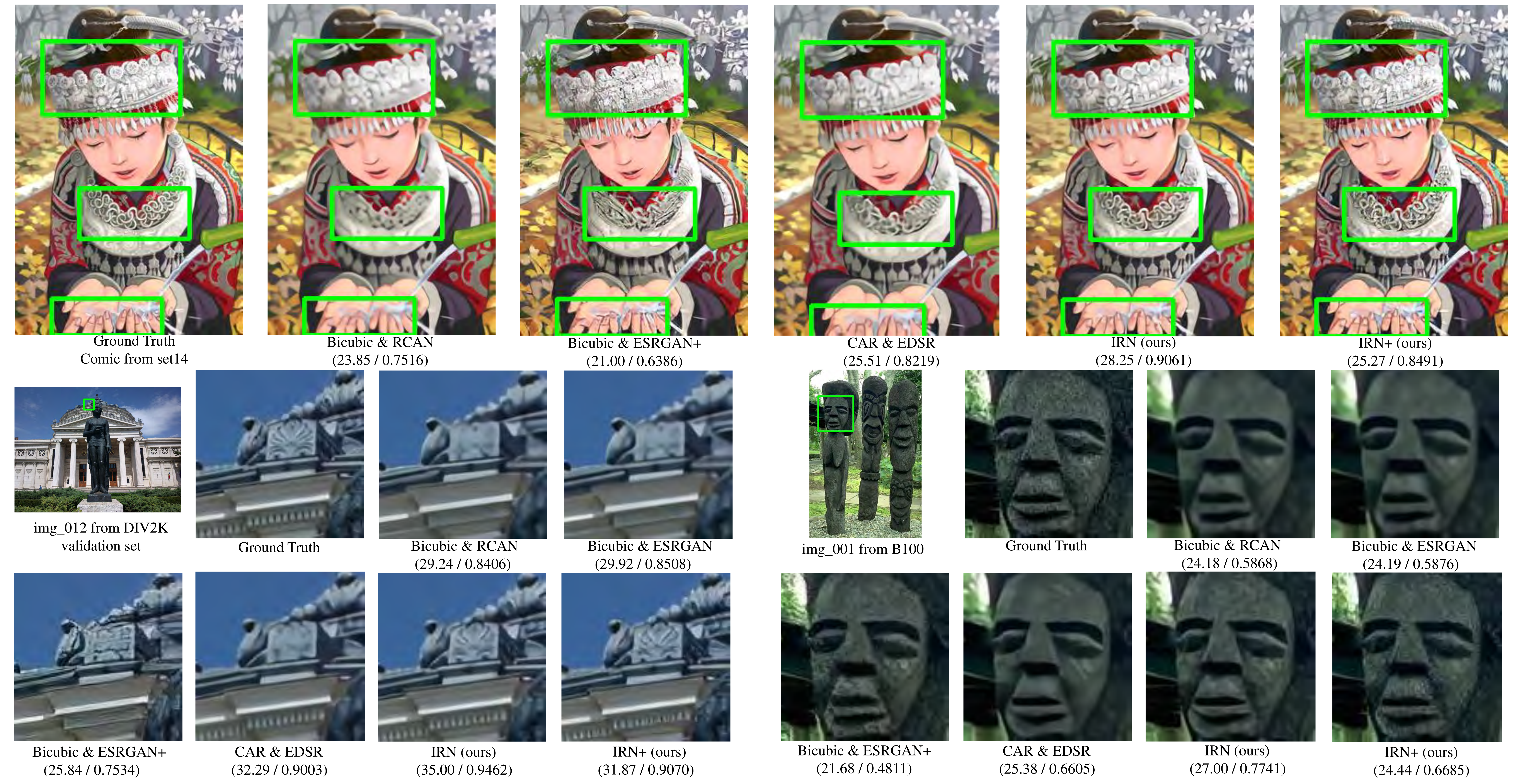}
    \caption{Qualitative results of upscaling the $4\times$ downscaled images. IRN recovers rich details, leading to both visually pleasing performance and high similarity to the original images. IRN+ produces even sharper and more realistic details. See the appendix for more results.}
    \label{fig:qualitative results}
    \vspace{-10pt}
\end{figure*}

\noindent\textbf{Qualitative Results}\quad
We then qualitatively evaluate IRN and IRN+ by demonstrating details of the upscaled images. As shown in Fig.~\ref{fig:qualitative results}, HR images reconstructed by IRN and IRN+ achieve better visual quality and fidelity than those of previous state-of-the-art methods. IRN recovers richer details, which contributes to the pleasing visual quality. IRN+ further produces sharper and more realistic images as the effect of the distribution matching objective. For the 'Comic' example, we observe that the IRN and IRN+ are the only models that can recover the complicated textures on the headwear and necklace, as well as the sharp and realistic fingers. Previous perceptual-driven methods such as ESRGAN~\cite{wang2018esrgan} also claim that the sharpness and reality of their generated HR images are satisfied. However, the visually unreasonable and unpleasing details produced by their model often lead to dissimilarity to the original images. We leave the high-resolution version and more results in the appendix for spacing reason.

\begin{figure} [ht]
\vspace{-10pt}
    \centering
    \subfigure[]{
    \includegraphics[scale=0.06]{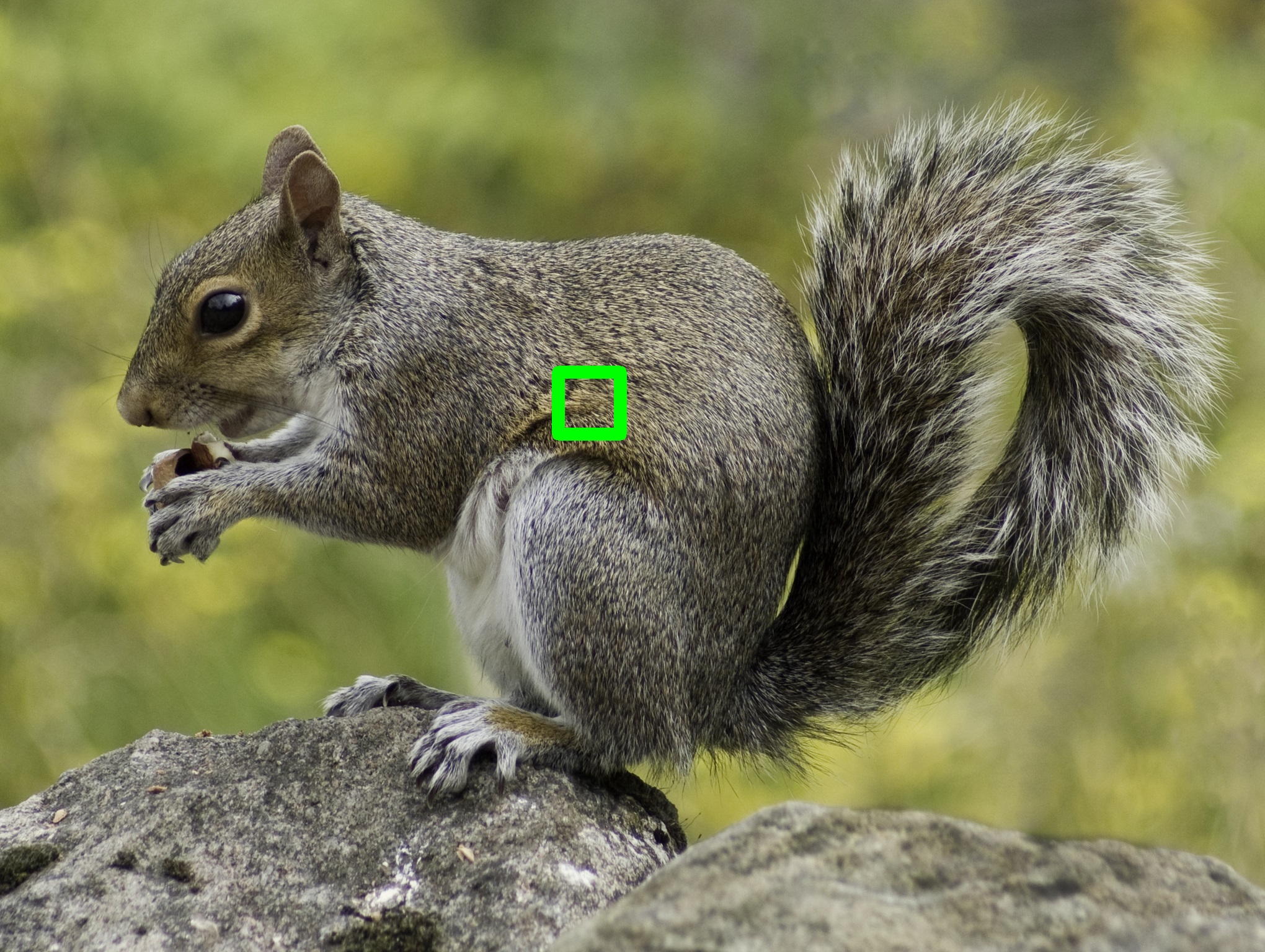}
    \label{ground truth}
    }
    \subfigure[]{
    \includegraphics[scale=0.9]{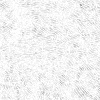}
    \label{detail1}
    }
    \subfigure[]{
    \includegraphics[scale=0.9]{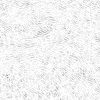}
    \label{detail2}
    }
    \subfigure[]{
    \includegraphics[scale=0.9]{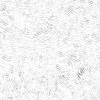}
    \label{detail3}
    }

    \caption{Visualisation of the difference of upscaled HR images from multiple draws of $z$. (a): original image;
    (b-d): HR image differences of three $z$ drawn from a common $z$ sample. Darker color means larger difference. It shows that the differences are random noise in high-frequency regions without a typical texture.
    }
    \label{fig:different samples}
    \vspace{-5pt}
\end{figure}

\begin{figure*} [ht]
    \centering
    \includegraphics[scale=0.3]{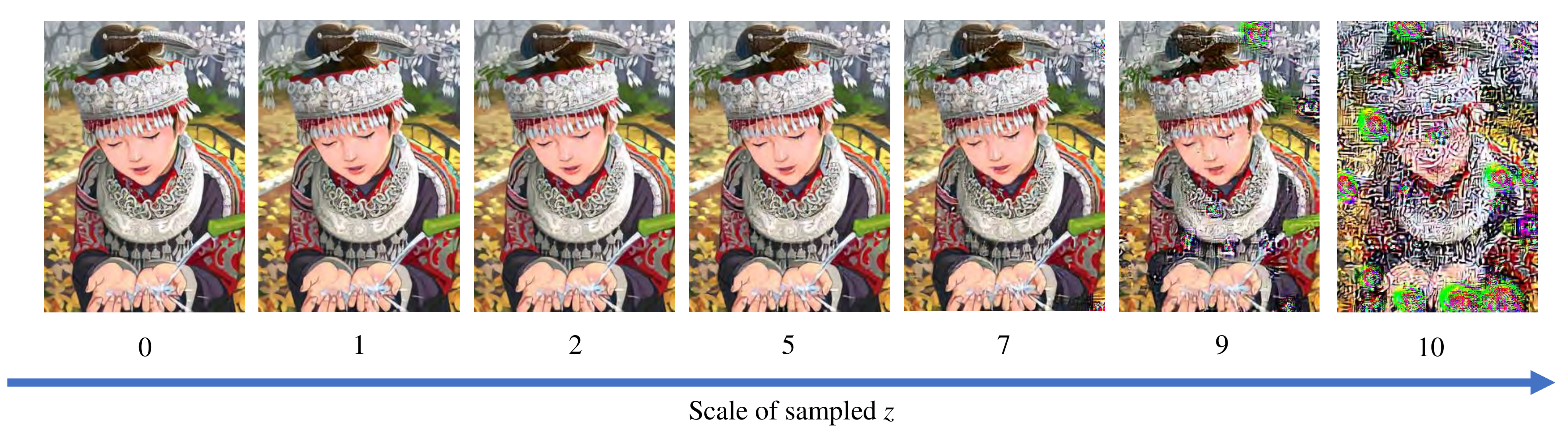}
    \caption{Results of HR images by IRN+ with out-of-distribution samples of $z$. We train $z$ with an isotropic Gaussian distribution, and illustrate upscaling results when scaling $z$ sampled from the isotropic Gaussian distribution.}
    \label{fig:scale results}
    \vspace{-15pt}
\end{figure*}

\noindent\textbf{Visualisation on the Influence of $z$}\quad
As described in previous sections, we aim to let $z\sim p(z)$ focus on the randomness of high-frequency contents only. In Table~\ref{quantitative results}, the PSNR difference is less than 0.02 dB for each image with different samples of $z$. In order to verify whether $z$ has learned only to influence high-frequency information, we calculate and present the difference between different draws of $z$ in Fig.~\ref{fig:different samples}. We can see in the figure that there is only a tiny noisy distinction in high-frequency regions without typical textures, which can hardly be perceived when combined with low-frequency contents. This indicates that our IRN has learned to reconstruct most meaningful high-frequency contents, while embedding senseless noise into randomness.

As mentioned above, we train the model to encourage $p(z)$ to obey a simple and easy-to-sample distribution, i.e., isotropic Gaussian distribution. In order to further verify the effectiveness of the learned model, we feed $(y, \alpha z)$ into our IRN+ to obtain $x_\alpha$ by controlling the scale of sampled $z$ with different values of $\alpha$.  As shown in Fig.~\ref{fig:scale results}, a larger deviation to the original distribution results in more noisy textures and distortion. It demonstrates that our model transforms $z$ faithfully to follow the specified distribution, and is also robust to slight distribution deviation.

\begin{table*} [ht]
\vspace{-10pt}
	\centering
	\caption{Analysis results (PSNR/SSIM) of training IRN with $L_1$ or $L_2$ LR guide and HR reconstruction loss, with/without partial distribution matching loss, on Set5, Set14, BSD100, Urban100 and DIV2K validation sets with scale 4$\times$.}
	\begin{tabular}{c|c|c|c|c|c|c|c}
		\hline
		$L_{guide}$ & $L_{recon}$ & $L_{distr'}$ & Set5 & Set14 & BSD100 & Urban100 & DIV2K \\
		\hline
		\hline
		$L_1$ & $L_1$ & Yes & 34.75 / 0.9296 & 31.42 / 0.8716 & 30.42 / 0.8451 & 30.11 / 0.8903 & 33.64 / 0.9079 \\
		\hline
		$L_1$ & $L_2$ & Yes & 34.93 / 0.9296 & 31.76 / 0.8776 & 31.01 / 0.8562 & 30.79 / 0.8986 & 34.11 / 0.9116 \\
		\hline
		$L_2$ & $L_1$ & Yes & \textcolor{red}{36.19} / 0.9451 & \textcolor{red}{32.67} / \textcolor{red}{0.9015} & \textcolor{red}{31.64} / \textcolor{red}{0.8826} & \textcolor{red}{31.41} / \textcolor{red}{0.9157} & \textcolor{red}{35.07} / \textcolor{red}{0.9318} \\
		\hline
		$L_2$ & $L_2$ & Yes & 35.93 / 0.9402 & 32.51 / 0.8937 & 31.64 / 0.8742 & 31.40 / 0.9105 & 34.90 / 0.9308 \\
		\hline
		\hline
		$L_2$ & $L_1$ & No &  36.12 / \textcolor{red}{0.9455} & 32.18 / 0.8995 & 31.49 / 0.8808 & 30.91 / 0.9102 & 34.90 / 0.9308 \\
		\hline
		
	\end{tabular}
	\label{ablation_loss}
	\vspace{-10pt}
\end{table*}

\noindent\textbf{Analysis on the Losses}\quad
We conduct experiments to analyze the components in the loss of Eqs.~(\ref{eq.guide loss}, \ref{eq.recon loss}, \ref{eq.distr loss'}). As shown in Table~\ref{ablation_loss}, IRN performs the best when the LR guidance loss is the $L_2$ loss and the HR reconstruction loss is the $L_1$ loss. The reason is that the $L_1$ loss encourages more pixel-wise similarity, while the $L_2$ loss is less sensitive to minor changes. In the forward procedure, we utilize the Bicubic-downscaled images as guidance, but we do not aim to exactly learn the Bicubic downscaling, which may harm the inverse procedure. The forward reconstruction loss only acts as a constraint to maintain visually pleasing downscaling, so the $L_2$ loss is more suitable. In the backward procedure, on the other hand, our goal is to reconstruct the ground truth image accurately. Therefore, the $L_1$ loss is more appropriate, as also identified by other super-resolution works.
Table~\ref{ablation_loss} also demonstrates the necessity of the partial distribution matching loss of Eq.~\eqref{eq.distr loss'}, which restricts the marginal distributions on $\clZ$, and benefits the forward distribution learning.

\vspace{-5pt}
\subsection{Evaluation on Downscaled LR Images}
\vspace{-2pt}

We also evaluate the quality of LR images downscaled by our IRN. We demonstrate the similarity index between our LR images and Bicubic-based LR images, and present similar visual perception of them, to show that IRN is able to perform as well as Bicubic.

\begin{table} [ht]
\vspace{-5pt}
	\centering
	\small
	\tabcolsep=2mm
	\caption{SSIM results between the images downscaled by IRN and by Bicubic on the Set5, Set14, BSD100, Urban100 and DIV2K validation sets.}
	\begin{tabular}{c|c|c|c|c|c}
		\hline
		Scale & Set5 & Set14 & BSD100 & Urban100 & DIV2K \\
		\hline
		\hline
		$2\times$ & 0.9957 & 0.9936 & 0.9936 & 0.9941 & 0.9945 \\
		\hline
		$4\times$ & 0.9964 & 0.9927 & 0.9923 & 0.9916 & 0.9933 \\
		\hline
		
	\end{tabular}
	\label{lr_ssim}
	\vspace{-5pt}
\end{table}

As shown in Table~\ref{lr_ssim}, images downscaled by IRN are extremely similar to those by Bicubic. Fig.~\ref{fig:downscaled images} and more figures in the appendix illustrate the visual similarity between them, which demonstrates the proper perception of our downscaled images.

\begin{figure} [htbp]
\vspace{-10pt}
    \centering
    \subfigure[]{
    \includegraphics[scale=0.6]{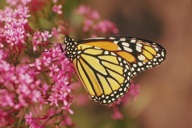}
    \label{img1_lrgt}
    }
    \subfigure[]{
    \includegraphics[scale=0.6]{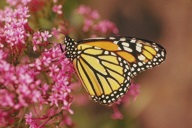}
    \label{img1_lr}
    }
    \subfigure[]{
    \includegraphics[scale=0.2]{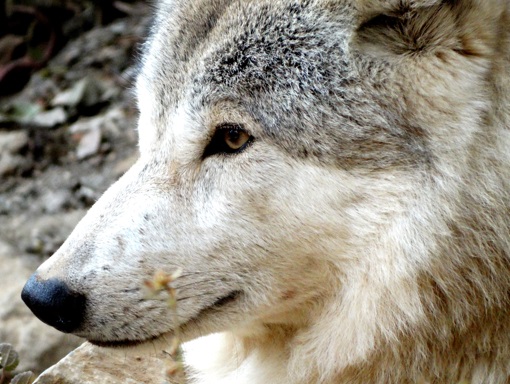}
    \label{img2_lrgt}
    }
    \subfigure[]{
    \includegraphics[scale=0.2]{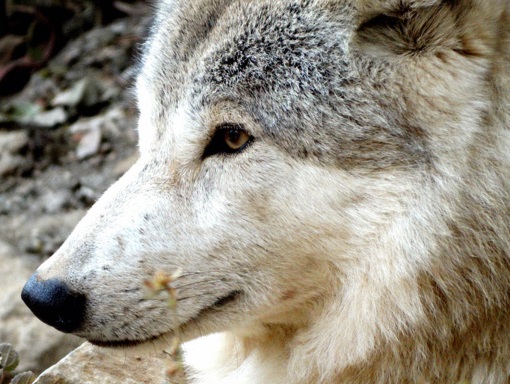}
    \label{img2_lr}
    }
    \caption{Demonstration of the downscaled images from Set14 and DIV2K validation sets. Left column (a,c): Image downscaled by Bicubic. Right column (b,d): Image downscaled by IRN. They share a similar visual perception.}
    \label{fig:downscaled images}
    \vspace{-15pt}
\end{figure}

\section{Conclusion}
\vspace{-2pt}


In this paper, we propose a novel invertible network for the image rescaling task, with which the ill-posed nature of the task is largely mitigated. We explicitly model the statistics of the case-specific high-frequency information that is lost in downscaling as a latent variable following a specified case-agnostic distribution which is easy to sample from. The network models the rescaling processes by invertibly transforming between an HR image and an LR image with the latent variable. With the statistical knowledge of the latent variable, we draw a sample of it for upscaling from a downscaled LR image (whose specific high-frequency information was lost during downscaling, of course). We design a specific invertible architecture tailored for image rescaling, and an effective training objective to enforce the model to have desired downscaling and upscaling behavior, as well as to output the latent variable with the specified properties. Extensive experiments demonstrate that our model significantly improves both quantitative and qualitative performance of upscaling reconstruction from downscaled LR images, while being light-weighted.

\clearpage

\bibliographystyle{splncs04}
\bibliography{IRN}

\newpage

\appendix
\begin{center}
\textbf{\Large Appendix: Invertible Image Rescaling}
\end{center}

\section{Details of the distribution loss}

According to the main text, we choose the Jensen-Shannon (JS) divergence as the distribution metric and minimize the difference between ${f_\theta^{-1}}_\# \big[ {f_\theta^y}_\# [q(x)] \; p(z) \big]$ and $q(x)$:

\begin{align}
	& \Ldistr(\theta) = \mathrm{JS}( {f_\theta^{-1}}_\# \big[ {f_\theta^y}_\# [q(x)] \; p(z) \big], q(x) ) \notag \\
	={} & \frac{1}{2} \max_T \Big\{ \bbE_{q(x)} \left[ \log \sigma(T(x)) \right] \notag \\
	    & {} + \bbE_{x' \sim {f_\theta^{-1}}_\# \big[ {f_\theta^y}_\# [q(x)] \; p(z) \big]} \left[ \log \left( 1 - \sigma(T(x')) \right) \right] \Big\} + \log 2 \notag \\
	={} & \frac{1}{2} \max_T \big\{ \bbE_{q(x)} \left[ \log \sigma(T(x)) \right] \notag \\
	    & {} + \bbE_{ (y, z) \sim {f_\theta^y}_\# [q(x)] \; p(z) } \left[ \log \left( 1 - \sigma(T( f_\theta^{-1} (y, z) )) \right) \right] \big\} + \log 2 \notag \\
	\approx{} & \frac{1}{2 N} \max_T \sum_n \Big\{ \log \sigma(T (x^{(n)})) \notag \\
	    & {} + \log \left( 1 - \sigma(T( f_\theta^{-1} ( f_\theta^y (x^{(n)}), z^{(n)} ) )) \right) \Big\} + \log 2.
\end{align}

The first equality stems from the variational form of the JS divergence which is composed for training generative adversarial nets~\cite{goodfellow2014generative}.
The second equality is a reformulation using the definition of pushed-forward distribution.
The third approximate equality leads to a Monte Carlo estimation to the objective function using the corresponding samples: $\{z^{(n)}\}_{n=1}^N$ i.i.d. drawn from $p(z)$, and $\{x^{(n)}\}_{n=1}^N \sim q(x)$.

\section{Detailed Training Strategies on DIV2K dataset}
We train and compare our model in $2\times$ and $4\times$ downscaling scale with one and two downscaling modules respectively. Each downscaling module has 8 InvBlocks and downscale the original image by $2\times$. We use Adam optimizer~\cite{kingma2014adam} with $\beta_1=0.9, \beta_2=0.999$ to train our model. The mini-batch size is set to 16. The input HR image is randomly cropped into $144 \times 144$ and augmented by applying random horizontal and vertical flips. In the pre-training stage, the total number of iteration is $50K$, and the learning rate is initialized as $2\times10^{-4}$ where halved at $[10k, 20k, 30k, 40k]$ mini-batch updates. The hyper-parameters in Eqn.10 are set as $\lambda_1=1,\lambda_2=16,\lambda_3=1$. After pre-training, we finetune our model for another $20K$ iterations as described in Sec.3.3. The learning rate is initialized as $1\times10^{-4}$ and halved at $[5k, 10k]$ iterations. We set $\lambda_1=0.01,\lambda_2=16, \lambda_3=1, \lambda_4=0.01$ in Eqn.11 and pre-train the discriminator for 5000 iterations. The discriminator is similar to \cite{ledig2017photo}, which contains eight convolutional layers with $3\times3$ kernels, whose numbers increase from 64 to 512 by a factor 2 each two layers, and two dense layers with 100 hidden units.

\section{Quantitive results of IRN+}

\vspace{-10pt}
\begin{table*} [ht]
	\centering
	\scriptsize
	\caption{Quantitative evaluation results (PSNR / SSIM) of different 4$\times$ image downscaling and upscaling methods on benchmark datasets: Set5, Set14, BSD100, Urban100, and DIV2K validation set. For our model, differences on average PSNR / SSIM of different samples for z are less than 0.02. We report the mean result. The best result is in red, while the second is in blue.}
	\begin{tabular}{c|c|c|c|c|c|c|c}
		
		\hline
		Downscaling \& Upscaling  &  Scale & Param  &  Set5  &  Set14  &  BSD100  &  Urban100  &  DIV2K \\
	
		\hline  
		\hline
	
		Bicubic \& Bicubic & 4$\times$ & / & 28.42 / 0.8104 & 26.00 / 0.7027 & 25.96 / 0.6675 & 23.14 / 0.6577 & 26.66 / 0.8521 \\

		\hline
		Bicubic \& SRCNN \cite{dong2015image} & 4$\times$ & 57.3K &  30.48 / 0.8628  &  27.50 / 0.7513  &  26.90 / 0.7101  &  24.52 / 0.7221 &   --  \\
		
		
		\hline
		Bicubic \& EDSR \cite{lim2017enhanced} & 4$\times$ & 43.1M & 32.62 / 0.8984 & 28.94 / 0.7901 & 27.79 / 0.7437 & 26.86 / 0.8080 & 29.38 / 0.9032 \\

		\hline
		Bicubic \& RDN \cite{zhang2018residual} & 4$\times$ & 22.3M &  32.47 / 0.8990  &  28.81 / 0.7871  &  27.72 / 0.7419  &  26.61 / 0.8028 &   --  \\
		
		\hline
		Bicubic \& RCAN \cite{zhang2018image} & 4$\times$ & 15.6M & 32.63 / 0.9002 & 28.87 / 0.7889 & 27.77 / 0.7436 & 26.82 / 0.8087 & 30.77 / 0.8460 \\
		
		\hline
		Bicubic \& ESRGAN \cite{wang2018esrgan} & 4$\times$ & 16.3M & 32.74 / 0.9012 & 29.00 / 0.7915 & 27.84 / 0.7455 & 27.03 / 0.8152 & 30.92 / 0.8486 \\
		
		\hline
		Bicubic \& SAN \cite{dai2019second} & 4$\times$ & 15.7M &  32.64 / 0.9003  &  28.92 / 0.7888  &  27.78 / 0.7436  &  26.79 / 0.8068  &  --  \\
		
		\hline
		TAD \& TAU \cite{kim2018task} & 4$\times$ & -- & 31.81 /  --  & 28.63 /  --   & 28.51 /  --   & 26.63 /  --   & 31.16 /  --   \\
		
		\hline
		CAR \& EDSR \cite{sun2020learned} & 4$\times$ & 52.8M & \textcolor{blue}{33.88} / \textcolor{blue}{0.9174} & \textcolor{blue}{30.31} / 0.8382 & \textcolor{blue}{29.15} / 0.8001 & \textcolor{blue}{29.28} / \textcolor{blue}{0.8711} & \textcolor{blue}{32.82} / 0.8837 \\
		
		\hline
		IRN (ours) & 4$\times$ & 4.35M & \textcolor{red}{36.19} / \textcolor{red}{0.9451} & \textcolor{red}{32.67} / \textcolor{red}{0.9015} & \textcolor{red}{31.64} / \textcolor{red}{0.8826} & \textcolor{red}{31.41} / \textcolor{red}{0.9157} & \textcolor{red}{35.07} / \textcolor{red}{0.9318} \\
        
		\hline
		IRN+ (ours) & 4$\times$ & 4.35M & 33.59 / 0.9147 & 29.97 / \textcolor{blue}{0.8444} & 28.94 / \textcolor{blue}{0.8189} & 28.24 / 0.8684 & 32.24 / \textcolor{blue}{0.8921} \\
		
        \hline
	    
	\end{tabular}
	
    \label{quantitative results of IRN+}
    
\end{table*}

IRN+ aims at producing more realistic images by minimizing the distribution difference, not exactly matching details of original images as IRN does. The difference will lead to lower PSNR and SSIM, which is the same as GAN-based super-resolution methods. Despite the difference, IRN+ still outperforms most methods in PSNR and SSIM as shown in Table.\ref{quantitative results of IRN+}, demonstrating the good similarity between the reconstructed images and original HR images.

\section{Different samples of $z$}
\begin{figure} [htbp]
    \centering
    \subfigure[]{
    \includegraphics[scale=0.12]{attachment/0810_GT_frame.jpg}
    \label{ground truth}
    }
    \subfigure[]{
    \includegraphics[scale=0.12]{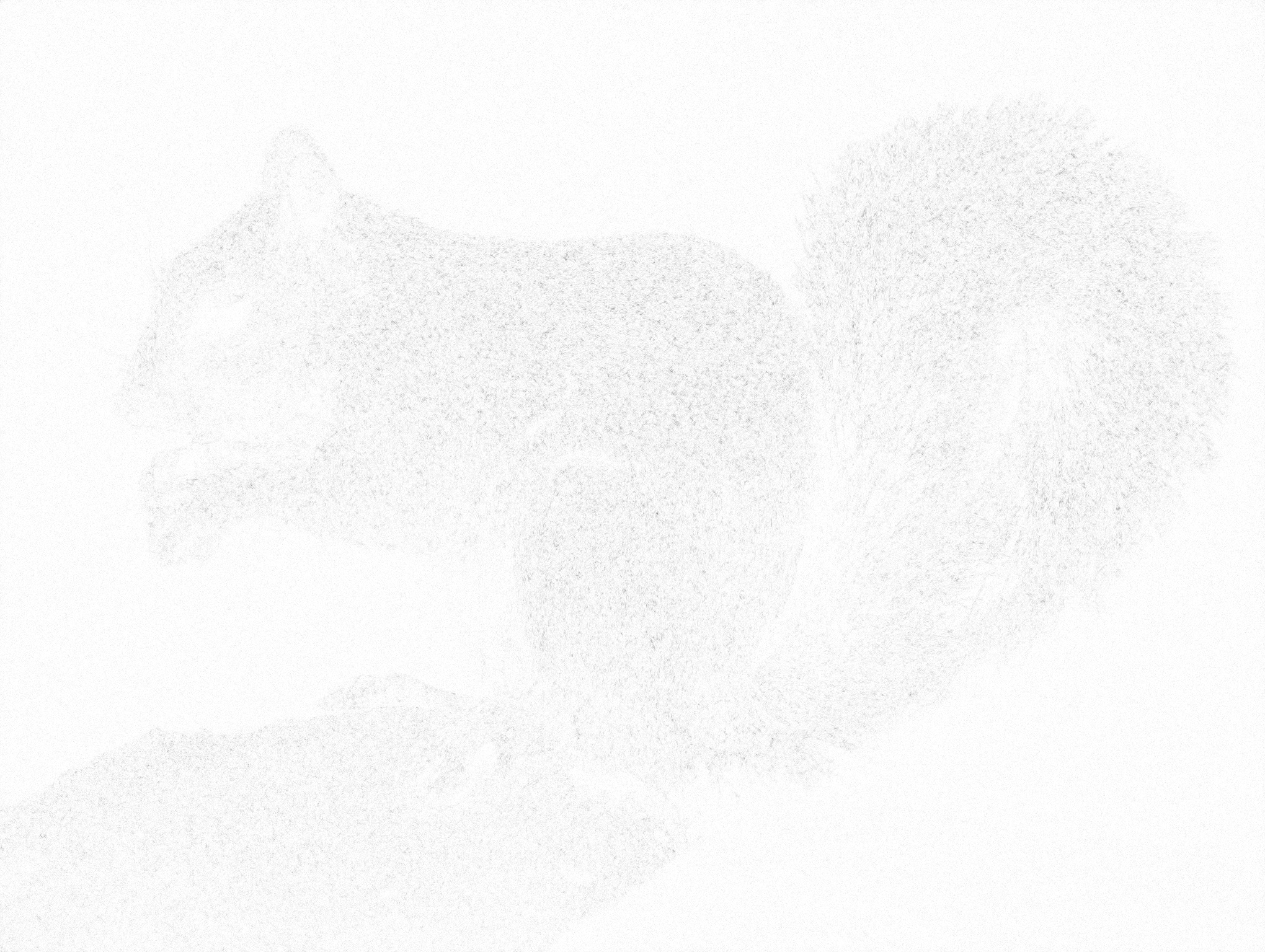}
    \label{sample0&sample1}
    }
    \subfigure[]{
    \includegraphics[scale=0.12]{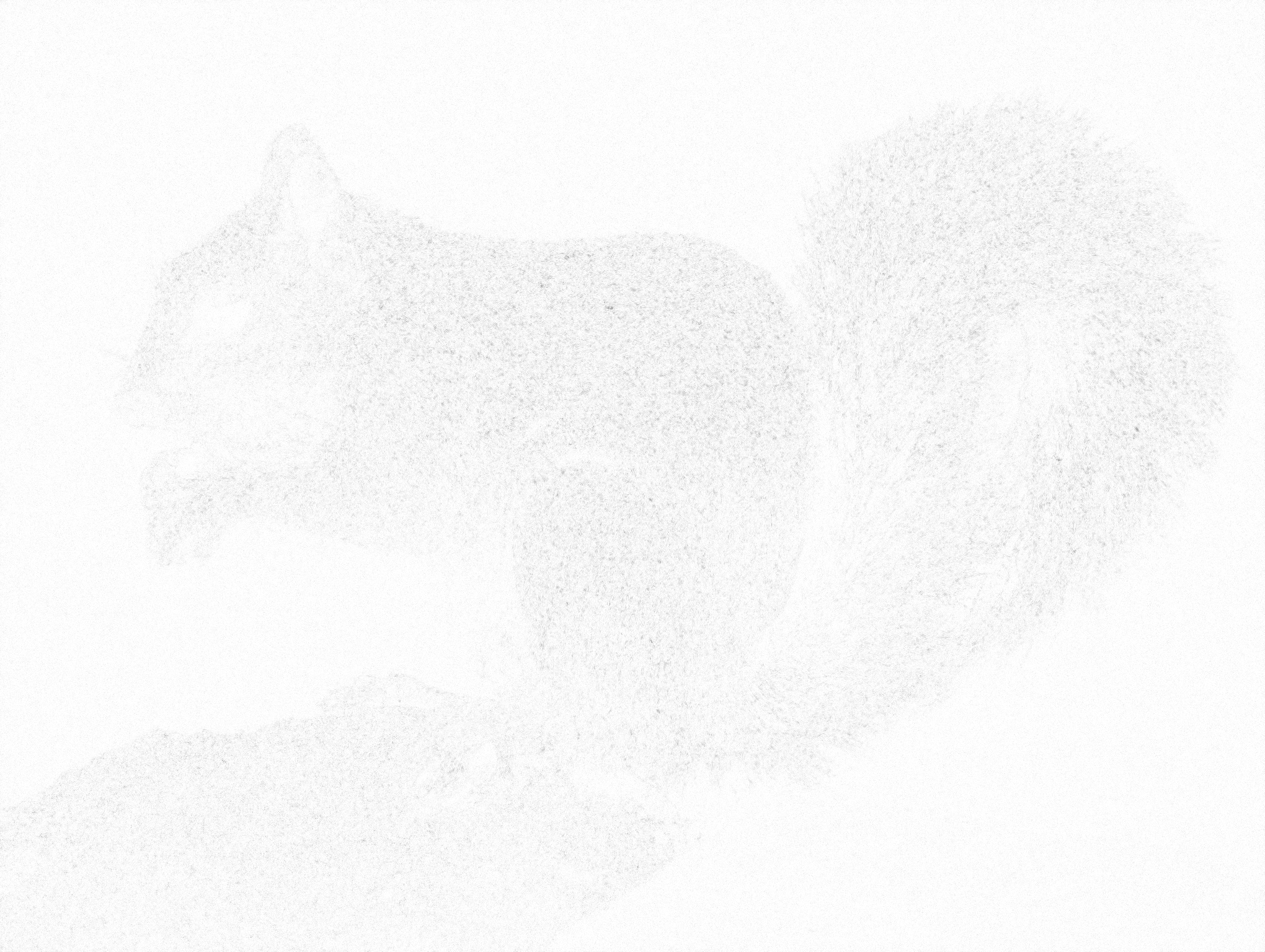}
    \label{sample0&sample2}
    }
    \subfigure[]{
    \includegraphics[scale=0.12]{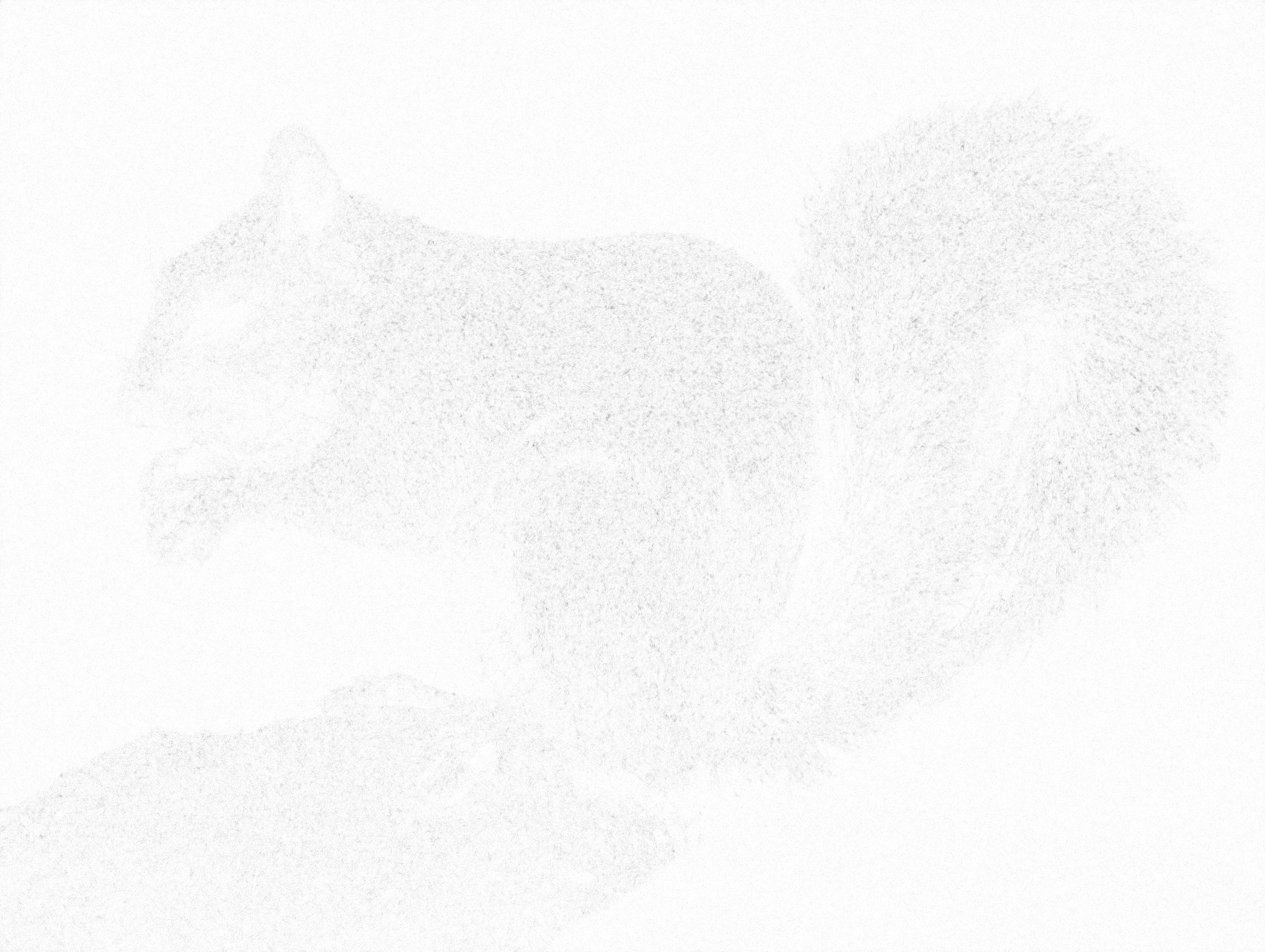}
    \label{sample0&sample3}
    }
    \subfigure[]{
    \includegraphics[scale=1.5]{attachment/grey_0810_diff01_patch_1.jpg}
    \label{detail1}
    }
    \hspace{2mm}
    \subfigure[]{
    \includegraphics[scale=1.5]{attachment/grey_0810_diff02_patch_1.jpg}
    \label{detail2}
    }
    \hspace{2mm}
    \subfigure[]{
    \includegraphics[scale=1.5]{attachment/grey_0810_diff03_patch_1.jpg}
    \label{detail3}
    }
    \caption{Difference between upscaled images by different samples of $z$. (a): Original image. (b-d): Residual of three randomly upscaled images with another sample (averaged over the three channels). (e-g): Detailed difference of (b-d). The darker the larger difference. To ensure the visual perception, we set rebalance factor by 20.
    }
    \label{fig:different samples}
\end{figure}

As shown in Fig.~\ref{fig:different samples}, there are only tiny noisy distinction in high-frequency areas without typical textures, which can hardly perceived when combined with low-frequency contents. Different samples lead to different but perceptually meaningless noisy distinctions.

\section{More qualitative results}

\begin{figure*} [htbp]
    \centering
    \includegraphics[scale=0.105]{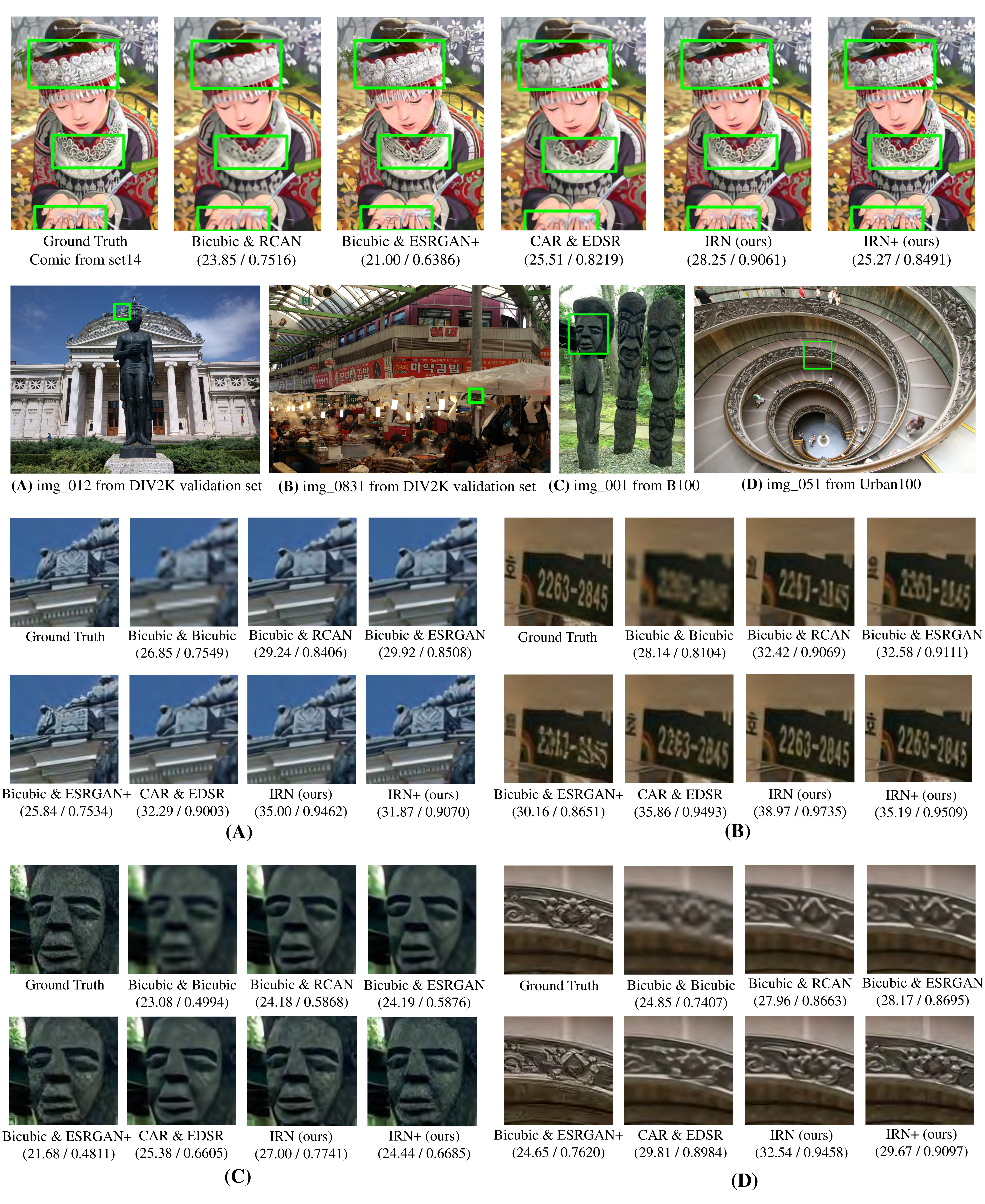}
    \caption{More qualitative results of upscaling the $4\times$ downscaled images on Set14, BSD100, Urban100 and DIV2K validation datasets.}
    \label{fig:qualitative results}
\end{figure*}

\begin{figure*} [htbp]
    \centering
    \includegraphics[scale=0.105]{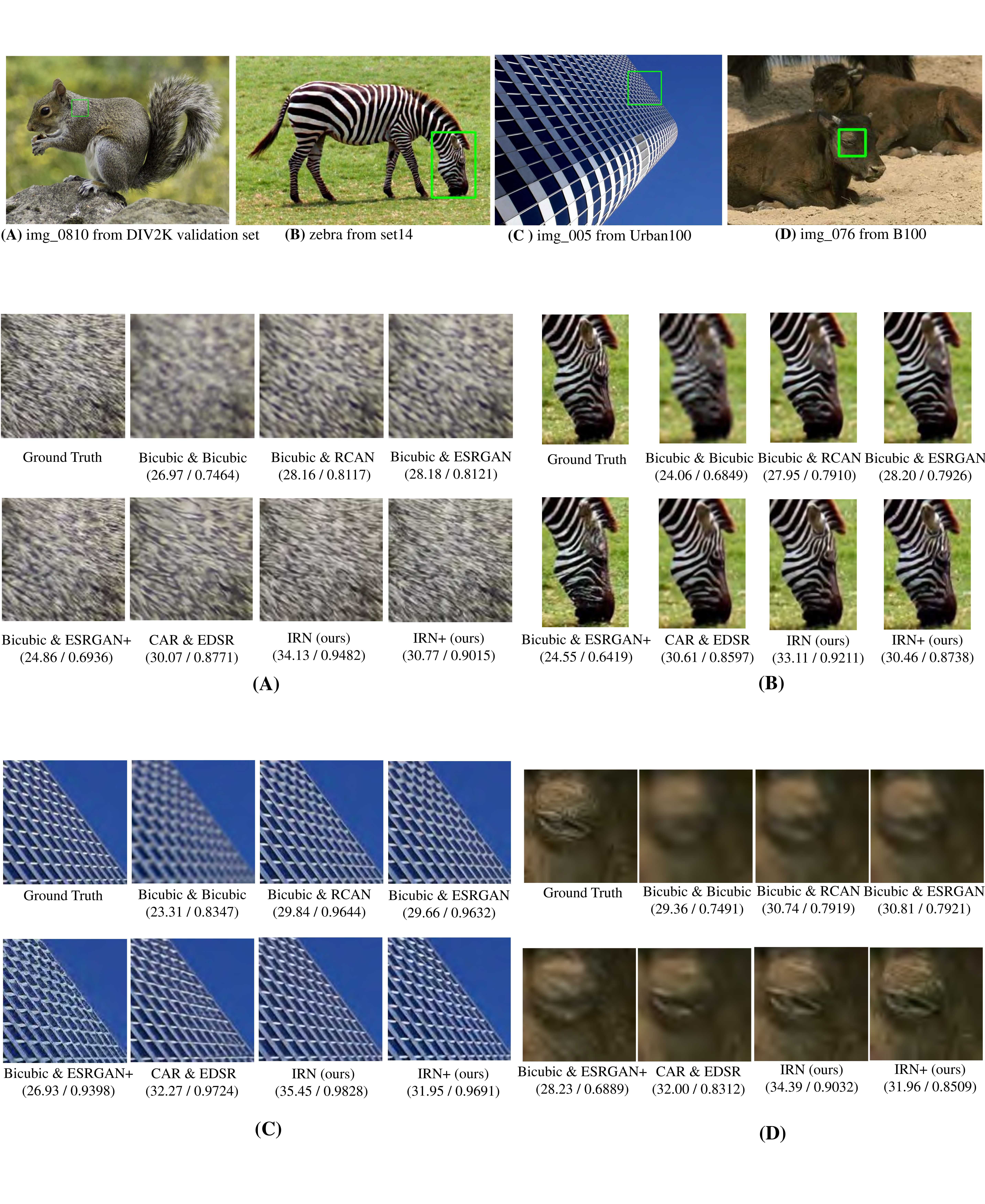}
    \caption{More qualitative results of upscaling the $4\times$ downscaled images on Set14, BSD100, Urban100 and DIV2K validation datasets.}
    \label{fig:qualitative results1}
\end{figure*}

\begin{figure*} [htbp]
    \centering
    \includegraphics[scale=0.105]{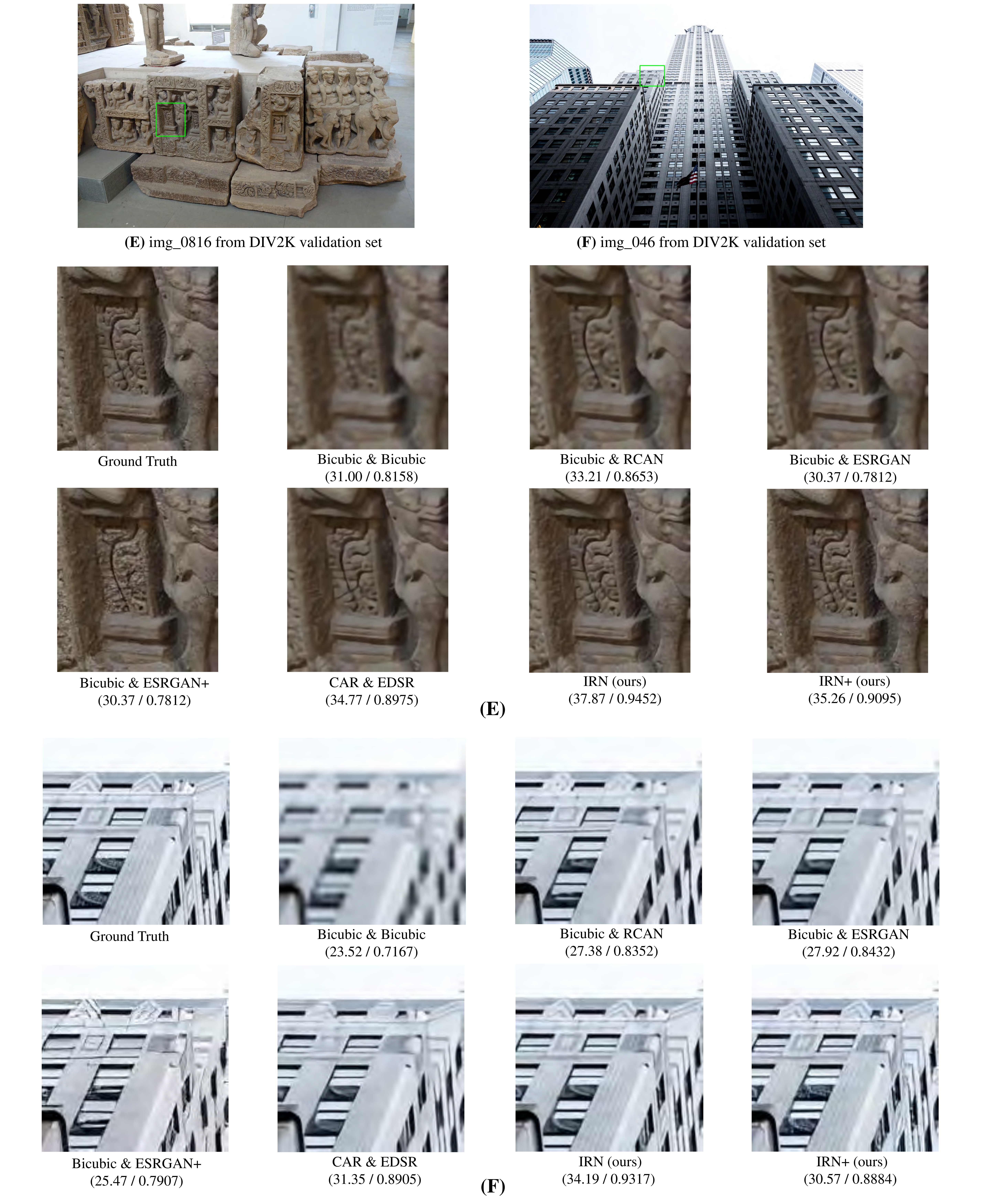}
    \caption{More qualitative results of upscaling the $4\times$ downscaled images on DIV2K validation dataset.}
    \label{fig:qualitative results2}
\end{figure*}

\begin{figure*} [htbp]
    \centering
    \includegraphics[scale=0.105]{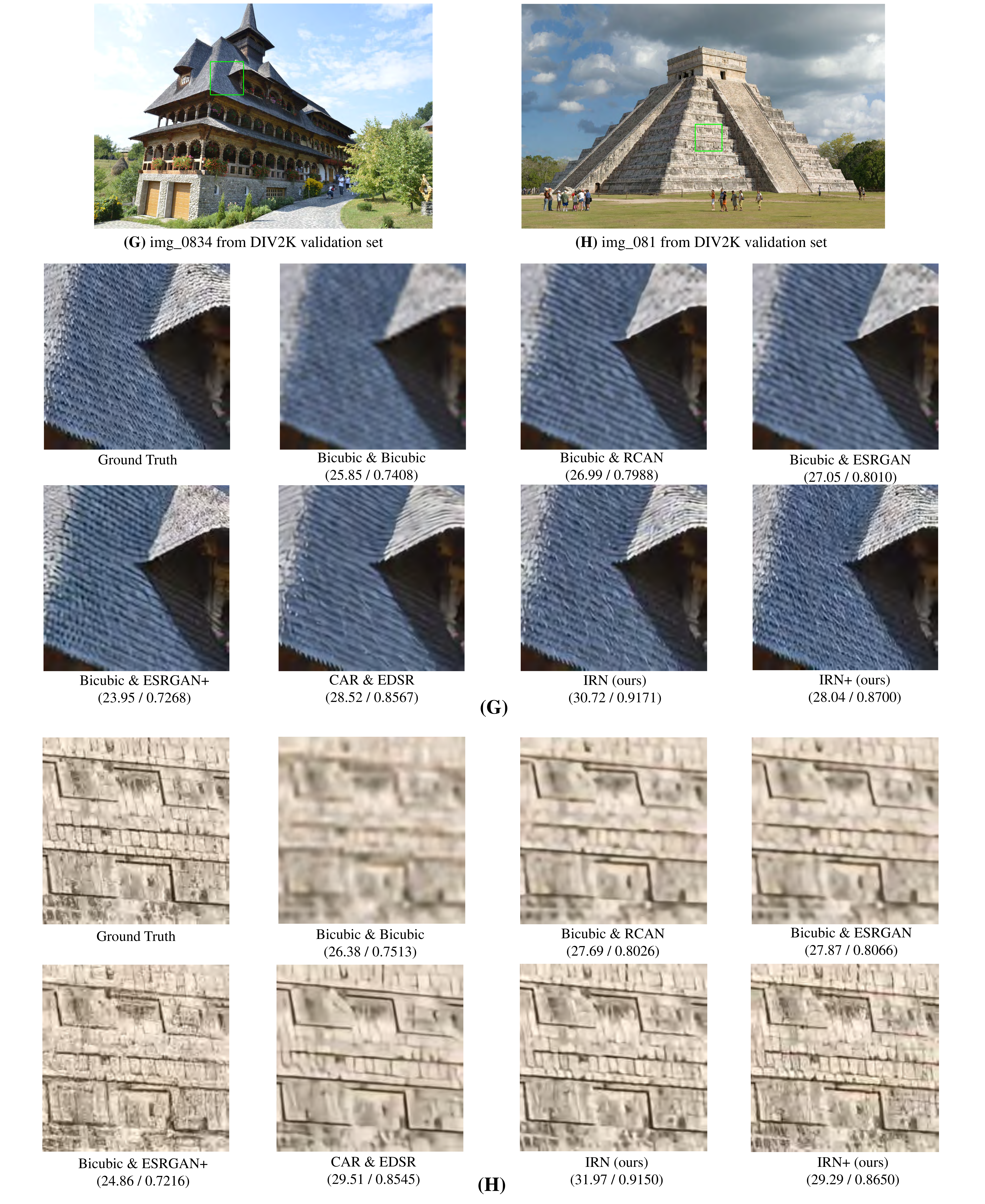}
    \caption{More qualitative results of upscaling the $4\times$ downscaled images on DIV2K validation dataset.}
    \label{fig:qualitative results3}
\end{figure*}

As shown in Fig.\ref{fig:qualitative results},\ref{fig:qualitative results1},\ref{fig:qualitative results2},\ref{fig:qualitative results3}, images reconstructed by IRN and IRN+ significantly outperforms previous both PSNR-oriented and perceptual-driven methods in visual quality and similarity to original images. IRN is able to reconstruct rich details including detailed lines and textures, which contributes to the pleasing perception. IRN+ further produce sharper and more realistic images as a result of the distribution matching objective.

\section{Evaluation on downscaled images}

As shown in Fig.~\ref{fig:downscaled images}, images downscaled by IRN share a similar visual perception with images downscaled by bicubic.

\vspace{-10pt}
\begin{figure} [htbp]
    \centering
    \subfigure[]{
    \includegraphics[scale=0.6]{attachment/img_011_SRF_4_HR_LRGT.jpg}
    \label{img1_lrgt}
    }
    \subfigure[]{
    \includegraphics[scale=0.6]{attachment/img_011_SRF_4_HR_LR.jpg}
    \label{img1_lr}
    }
    \subfigure[]{
    \includegraphics[scale=0.225]{attachment/0805_LRGT.jpg}
    \label{img2_lrgt}
    }
    \subfigure[]{
    \includegraphics[scale=0.225]{attachment/0805_LR.jpg}
    \label{img2_lr}
    }
    \\
    \subfigure[]{
    \includegraphics[scale=0.45]{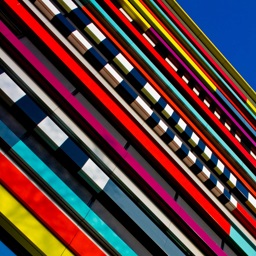}
    \label{img3_lrgt}
    }
    \subfigure[]{
    \includegraphics[scale=0.45]{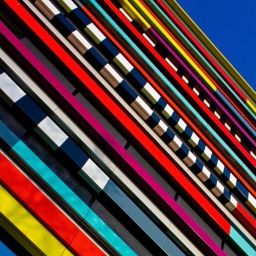}
    \label{img3_lr}
    }
    \subfigure[]{
    \includegraphics[scale=0.225]{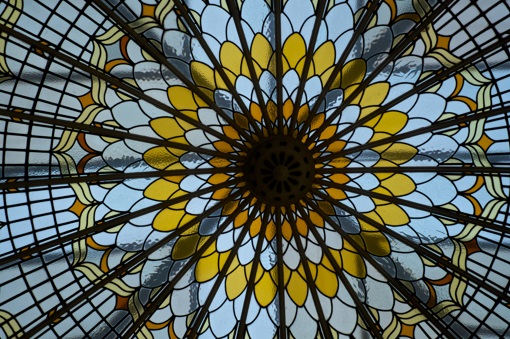}
    \label{img4_lrgt}
    }
    \subfigure[]{
    \includegraphics[scale=0.225]{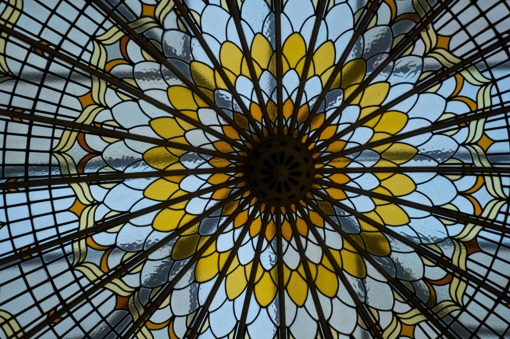}
    \label{img4_lr}
    }
    \\
    \subfigure[]{
    \includegraphics[scale=1.35]{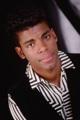}
    \label{img5_lrgt}
    }
    \subfigure[]{
    \includegraphics[scale=1.35]{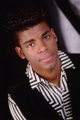}
    \label{img5_lr}
    }
    \subfigure[]{
    \includegraphics[scale=0.975]{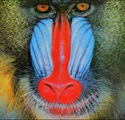}
    \label{img6_lrgt}
    }
    \subfigure[]{
    \includegraphics[scale=0.975]{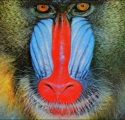}
    \label{img6_lr}
    }
    \\
    \subfigure[]{
    \includegraphics[scale=0.48]{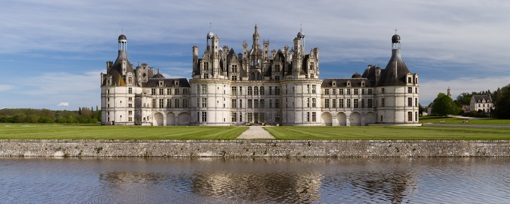}
    \label{img7_lrgt}
    }
    \subfigure[]{
    \includegraphics[scale=0.48]{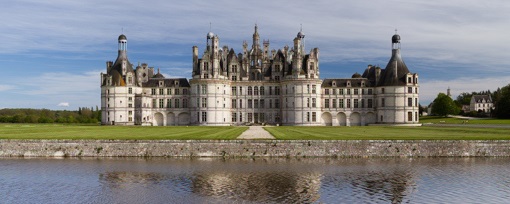}
    \label{img7_lr}
    }
    \caption{Demonstration of the downscaled images from Set14, B100, Urban100, and DIV2K validation set. Left column (a,c,e,g,i,k,m): Image downscaled by Bicubic. Right column (b,d,f,h,j,l,n): Image downscaled by IRN. They share a similar visual perception.}
    \label{fig:downscaled images}
\end{figure}

\end{document}